%% file: paper.tex
\newcommand*{\ATLASLATEXPATH}{latex/}
\documentclass[cernpreprint,texlive=2011,txfonts,UKenglish]{latex/atlasdoc} 
\pdfoutput=1
\usepackage{\ATLASLATEXPATH atlaspackage}
\usepackage{\ATLASLATEXPATH atlasbiblatex}
\usepackage{\ATLASLATEXPATH atlascontribute}
\usepackage{\ATLASLATEXPATH atlasphysics}
\graphicspath{{logos/}{figures/}}
\usepackage{paper-defs}
\input{paper-metadata}

\begin{document}
\maketitle

\tableofcontents

\section{Introduction}
\label{sec:intro}

Additional partners of the top quark are ingredients in several models that address the hierarchy 
problem~\cite{Weinberg:1975gm,Gildener:1976ai,Weinberg:1979bn,Susskind:1978ms} 
of the Standard Model (SM).  Supersymmetry (SUSY)~\cite{Miyazawa:1966,Ramond:1971gb,Golfand:1971iw,Neveu:1971rx,Neveu:1971iv,Gervais:1971ji,Volkov:1973ix,Wess:1973kz,Wess:1974tw} is one such model which naturally resolves the hierarchy problem with the introduction of supersymmetric partners of the known bosons and fermions.  A supersymmetric partner of the top quark would stabilise the Higgs boson mass against quadratically divergent quantum corrections, provided that its mass is close to the electroweak symmetry breaking energy scale. This would make its discovery possible at the Large Hadron Collider (LHC)~\cite{LHC}. In a generic $R$-parity-conserving minimal supersymmetric extension of the SM (MSSM)~\cite{Fayet:1976et,Fayet:1977yc,Farrar:1978xj,Fayet:1979sa,Dimopoulos:1981zb}, the scalar partners of right-handed and left-handed quarks, $\sqR$ and $\sqL$, can mix, as can the scalar partners of charged leptons, $\slR$ and $\slL$, to form two squark or two slepton mass eigenstates, respectively. The lighter of the two top squark eigenstates is denoted $\tone$ and is referred to as the \scalartop in the following. Likewise, the lighter of the two scalar tau eigenstates is denoted $\stauOne$ and referred to herein as the \scalartau.

In gauge-mediated supersymmetry breaking (GMSB) models~\cite{Dine:1981gu,AlvarezGaume:1981wy,Nappi:1982hm,Dine:1993yw, Dine:1994vc,Dine:1995ag}, the spin-3/2 partner of the graviton, called the gravitino $\tilde G$, is assumed to be the lightest  
supersymmetric particle. 
Assuming that the mass scale of the messengers responsible for the supersymmetry breaking is of the order of 10~\TeV, in order to minimise fine tuning~\cite{Barbieri:1987fn}, the  \scalartop should be lighter than about 400~\GeV~\cite{Asano:2010ut}. If the \scalartau is lighter  than the \scalartop, and the supersymmetric partners of the gauge and Higgs bosons (charginos and neutralinos) are heavier, the dominant decay mode of the $\tone$ might be the three-body decay into $b \nu_\tau \stauOne$, where $\nu_\tau$
is the tau neutrino, followed by the $\stauOne$ decay into a tau lepton and a gravitino. The other possible decay mode is the two-body decay into a top quark and a gravitino. 
The partial width of the two-body decay depends on the gravitino mass, while the partial width of the three-body decay via a virtual chargino depends on the chargino mass, as well as the chargino and scalar top mixing. For fixed scalar top and scalar tau masses either mode can dominate, and we focus in this paper on the signature resulting from the three-body decay. The two-body decay would give a signature very similar to that of the decay into a top quark and a neutralino, which has been addressed in previous searches~\cite{Aad:2015pfx,Chatrchyan:2012uea,Chatrchyan:2013lya,Chatrchyan:2013xna,Khachatryan:2015pwa, Khachatryan:2015vra, Khachatryan:2015wza}. In the simplest gauge-mediated models, the predicted Higgs boson mass~\cite{Arbey:2011ab} is typically lower than the measured mass~\cite{Aad:2015zhl}, especially if a light scalar top is also required. However, a variety of mechanisms exist~\cite{Craig:2011yk, Abazov:2008kz, Csaki:2012fh, Larsen:2012rq, Craig:2012hc} to raise the Higgs boson mass to make it  compatible with the observed value.

A lower limit of 87~\GeV\ on the mass of the scalar tau has been set by the LEP experiments~\cite{LEPcomb, Abbiendi:2005gc, Heister:2001nk, Abdallah:2003xe, Achard:2003ge}. 
No limits have been published so far from hadron collider searches for the three-body decay of the scalar top into the \scalartau. Searches for \scalartop pair production in proton--proton ($pp$) collisions, targeting the decay into charginos or neutralinos, have been performed by the ATLAS~\cite{Aad:2015pfx}
and CMS~\cite{Chatrchyan:2012uea,Chatrchyan:2013lya,Chatrchyan:2013xna,Khachatryan:2015pwa, Khachatryan:2015vra, Khachatryan:2015wza} collaborations. Searches for {\scalartop}s decaying into gravitinos, but not including the \scalartau in the decay chain, have
been reported by the ATLAS~\cite{Aad:2014mha} and CMS~\cite{CMS-SUS-13-014,CMS-SUS-13-024} collaborations.

\begin{figure}[htb!]
\begin{center}
\includegraphics[width=0.4\textwidth]{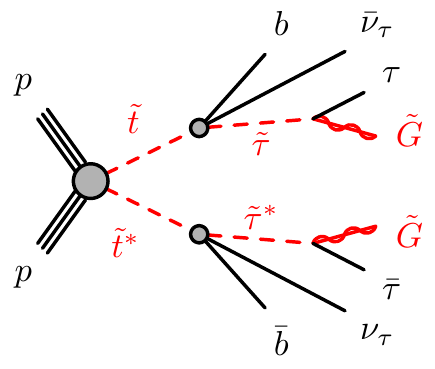}
\end{center}
\caption{Diagram showing the decay topology of the signal process.} 
\label{fig:diagram}
\end{figure}

This paper presents a dedicated search for pair production of scalar tops resulting in a final state with two tau leptons, two jets that contain a $b$--hadron ($b$-jets), and two very light gravitationally interacting particles. The decay topology of the signal process is shown in Fig.~\ref{fig:diagram}; the model considered is a simplified model in which all the supersymmetric particles other than the \scalartop and the ones entering its decay chain are decoupled. In order to maximise the sensitivity, two distinct analyses have been performed based on the decay mode of the tau leptons in the final state: one analysis requires two hadronically decaying tau leptons (the hadron--hadron channel) and the other requires one hadronically decaying tau lepton and one tau decaying into an electron or muon, plus neutrinos (the lepton--hadron channel). In addition, the results of the search reported in Ref.~\cite{Aad:2014qaa}, which is sensitive to events where both tau leptons decay leptonically (referred to as the lepton--lepton channel), are reinterpreted and limits are set on the \scalartop and scalar tau masses.

\section{ATLAS detector}
\label{sec:detector}

ATLAS~\cite{atlas-detector} is a multi-purpose particle physics experiment at the LHC.
The ATLAS detector\footnote{ATLAS uses a right-handed coordinate system with its origin at the nominal interaction point (IP) in the centre of the detector and the $z$-axis coinciding with the axis of the beam pipe. The $x$-axis points from the IP to the centre of the LHC ring, and the $y$-axis points upwards. Cylindrical coordinates
$(r,\phi)$ are used in the transverse plane, $\phi$ being the azimuthal angle around the beam pipe. The pseudorapidity is defined in terms of the polar angle $\theta$ as $\eta = -\ln \tan(\theta/2)$.} consists of an inner tracking detector surrounded by a superconducting solenoid, electromagnetic and hadronic calorimeters, and a muon spectrometer. 
The inner detector covers $| \eta | < 2.5$ and consists of a silicon pixel detector, a semiconductor microstrip detector, and a transition radiation tracker (TRT). The inner detector is surrounded by a thin superconducting solenoid providing a 2 T axial magnetic field, and allows for precision tracking of charged particles and vertex reconstruction. The calorimeter system covers the pseudorapidity range $| \eta | < 4.9$.
In the region  $| \eta | < 3.2$, high-granularity liquid-argon electromagnetic sampling calorimeters are used. A steel/scintillator-tile calorimeter provides energy measurements for hadrons within  $| \eta | < 1.7$. The end-cap and forward regions, which cover the range $1.5 < | \eta | < 4.9$, are instrumented with liquid-argon calorimeters for electromagnetic and hadronic particles.
The muon spectrometer surrounds the calorimeters and consists of three large superconducting air-core toroid magnets, each with eight coils, 
a system of tracking chambers (covering $|\eta| < 2.7$) and fast trigger chambers (covering $| \eta | < 2.4$).

\section{Monte Carlo simulations and data samples}

\label{sec:MC}
A number of Monte Carlo (MC) simulated event samples are used to model the signal and describe the backgrounds. 
For the main background components, predictions are normalised to the data in control regions (CRs) 
and then extrapolated to the signal regions (SRs) using simulation. All MC samples utilised in the
analyses are processed using either the ATLAS detector simulation~\cite{Aad:2010ah} based on {\tt GEANT4}~\cite{Agostinelli:2002hh} or
 a fast simulation based on a parameterisation of the performance of the
ATLAS electromagnetic and hadronic calorimeters~\cite{ATLAS-PUB-atlfast} and {\tt GEANT4} elsewhere.  Additional $pp$ interactions 
in the same (in-time) and nearby (out-of-time) bunch crossings, termed pile-up, are included in the simulation, and events are
reweighted so that the distribution of the number of pile-up collisions matches that in the data.

The signal model considered is a supersymmetric model with the gravitino as the lightest supersymmetric particle. 
By construction, the scalar partner of the right-handed tau lepton and the lightest scalar top\footnote{The mixing matrix of the simulated samples is such that the lightest scalar top eigenstate is almost a pure partner of the right-handed top quark.} are the next-to-lightest and 
the next-to-next-to-lightest supersymmetric particles, respectively, and different signal models are simulated by varying their masses. 
Pair production of the \scalartop is generated using {\tt HERWIG++ 2.6.3}~\cite{Bahr:2008pv} with the parton distribution functions (PDF) set CTEQ6L1~\cite{Nadolsky:2008zw}. 
The model requires that the \scalartop decays to $b \nu_\tau \stauOne$ via a virtual chargino with 100\% branching ratio, while
the $\stauOne$ decays, with a 100\% branching ratio, into a tau lepton and a gravitino. Lifetimes are assumed to be small enough 
(below about 1 ps) that the detector response is unaffected by the decay distance of the supersymmetric particles from the 
primary vertex. 

 Signal cross sections are calculated to next-to-leading order (NLO) in the strong coupling constant $\alpha_s$, adding the resummation of soft gluon emission at next-to-leading-logarithmic accuracy (NLO+NLL)~\cite{Beenakker:1997ut,Beenakker:2010nq,Beenakker:2011fu}. The nominal cross section and its uncertainty are taken from an envelope of cross-section predictions using different PDF sets and factorisation and renormalisation scales, as described in Ref.~\cite{Kramer:2012bx}.

\begin{table}
\begin{center}
\begin{small}
\caption{Details about the MC generation of the background and signal samples. \label{tab:mc}}
\begin{tabular}{L{0.19\textwidth}   C{0.165\textwidth}  C{0.165\textwidth}  C{0.135\textwidth}  C{0.10\textwidth}  C{0.1\textwidth} }
\hline

Process     & Generator   & Parton shower & Cross-section normalisation & PDF set & Generator tune \\ 
\hline

$t \bar t$    & {\tt POWHEG-BOX} r2129~\cite{Alioli:2010xd,powhegboxttbar}  & {\tt PYTHIA} 6.426  \cite{Sjostrand:2007gs} &  NNLO+NNLL \cite{Cacciari:2011hy,Baernreuther:2012ws,Czakon:2012zr,Czakon:2012pz,Czakon:2013goa,Czakon:2011xx}  & NLO CT10 \cite{Gao:2013xoa} & Perugia  2011C~\cite{Cooper:2011gk} \\ 

Single-top  ($Wt$~and~$s$-channel)& {\tt POWHEG-BOX}  r1556~\cite{Alioli:2010xd,Re:2010bp,Alioli:2009je}  &  {\tt PYTHIA} 6.426 & NNLO+NNLL \cite{Kidonakis:2010ux}   & CTEQ6L1 \cite{Nadolsky:2008zw}  & Perugia 2011C  \\ 

Single-top ($t$-channel) & {\tt ACERMC} 3.8 \cite{Kersevan:2004yg}  &  {\tt PYTHIA} 6.426 & NNLO+NNLL  \cite{Kidonakis:2011wy} & CTEQ6L1 & Perugia 2011C \\ 

\ttbar + $W/Z$ & {\tt MADGRAPH5} 1.3.28~\cite{Alwall:2011uj} & {\tt PYTHIA} 6.426 & NLO~\cite{Garzelli:2012bn} & CTEQ6L1 & AUET2 \cite{ATL-PHYS-PUB-2011-008} \\ 
 
$WW$, $WZ$, $ZZ$ & {\tt SHERPA} 1.4.1~\cite{Gleisberg:2008ta} &   {\tt SHERPA} 1.4.1 & NLO~\cite{Binoth:2006mf} & NLO CT10 & {\tt SHERPA} default \\ 

 $Z/\gamma^* (\rightarrow ee/\mu\mu)$+jets   & {\tt ALPGEN} 2.14~\cite{Mangano:2002ea} & {\tt HERWIG} 6.520 \cite{Corcella:2000bw} & NNLO \cite{dynnlo} & CTEQ6L1 & AUET2 \\ 

$Z/\gamma^* (\rightarrow \tau \tau)$+jets  & {\tt SHERPA} 1.4.1 & {\tt SHERPA} 1.4.1 & NNLO \cite{dynnlo} & NLO CT10 &  {\tt SHERPA} default \\  

$W(\rightarrow \ell \nu)$+jets, $\ell=e,\mu,\tau$  & {\tt SHERPA} 1.4.1  & {\tt SHERPA} 1.4.1 & NNLO \cite{dynnlo} & NLO CT10 &  {\tt SHERPA} default \\ 
 
 $\tone \tone^{*}$ & {\tt HERWIG++ 2.6.3} \cite{Bahr:2008pv} &  {\tt HERWIG++ 2.6.3} & NLO+NLL \cite{Beenakker:1997ut,Beenakker:2010nq,Beenakker:2011fu} & CTEQ6L1 & UE-EE-3 \cite{Gieseke:2012ft} \\ 
\hline
\end{tabular}
\end{small}
\end{center}
\end{table}

The programs used to generate signal and background events, as well as details of the cross-section calculation, PDF sets, and generator tunings, are reported
in  Table~\ref{tab:mc}. 

The data sample used in this paper was recorded between March and December 2012, with the LHC operating at a centre-of-mass energy of $\sqrt{s}=8$~TeV. The data are collected based on the decisions of a three-level trigger system~\cite{PERF-2011-02}. 
Events are selected for the electron--hadron channel if they are accepted by a single-electron trigger, and for the muon--hadron channel if accepted by a single-muon trigger. For the hadron--hadron channel, a missing transverse momentum trigger is used.
The trigger efficiency reaches its maximum value for leptons with a transverse momentum~($\pT$) above 25~\GeV\ in the lepton--hadron channels, and it exceeds 97\% for a missing transverse momentum above 150~\GeV\ in the hadron--hadron channel.
After beam, detector and data-quality requirements, the integrated luminosity of the data samples is $20.3 \,\ifb $ in the electron--hadron 
and muon--hadron channels, and $20.1 \,\ifb $~\cite{lumi} in the hadron--hadron channel. 
The difference in integrated luminosity is due to the additional data-quality requirements related to the trigger used in the hadron--hadron channel. 

\section{Event reconstruction}
\label{sec:obj}

The reconstruction and selection of final-state objects used in this analysis are discussed below.

Vertex candidates from $pp$ interactions are reconstructed using tracks in the inner detector. 
To identify the hard-scattering vertex in the presence of pile-up, the vertex with the highest scalar sum of the squared transverse momentum of the associated tracks, $\Sigma p^2_\mathrm{T}$, is defined as the primary vertex. The primary vertex is required to have at least five associated tracks
with $\pT > 400$~\MeV. 

Jets are reconstructed from three-dimensional clusters of energy deposits in the calorimeters using the anti-$k_t$ jet clustering algorithm~\cite{Cacciari:2008gp} using {\tt FastJet}~\cite{Cacciari:2005hq}, with a radius parameter of $R=0.4$. The differences in the calorimeter response between electrons/photons and 
hadrons are taken into account by classifying each cluster as coming from a hadronic or an electromagnetic shower on the basis
of its shape~\cite{Aad:2009wy}. The energy of electromagnetic and hadronic clusters is then weighted with correction factors derived from 
MC simulations. The average expected contribution from pile-up, calculated as the product of the jet area and the median energy density of the event~\cite{Cacciari:2008gn}, is subtracted from the jet energy. 
A further energy and $\eta$ calibration based on MC simulations and data, relating the 
response of the calorimeter to the true simulated jet energy~\cite{Aad:2011he, Aad:2012vm}, is then applied.
The jets selected in the analysis are the jet candidates with $\pT> 20 \GeV$ and $|\eta| < 2.5$.
Events containing jets that are likely to have arisen from detector noise, beam background or cosmic rays, are removed using the procedures described 
in Ref.~\cite{Aad:2011he}.   Events containing any jet failing to meet specific quality criteria described in Ref.~\cite{Aad:2013zwa} are also rejected.

Among the jets satisfying the selection criteria above, $b$-jet candidates are identified by a neural-network-based algorithm, which utilises the impact parameters of tracks, secondary vertex reconstruction, and the topology of $b$- and $c$-hadron decays inside a jet~\cite{btag1,btag2}. The efficiency for tagging $b$-jets in a MC sample of $t\bar{t}$ events using this algorithm is 70\% with rejection factors of 137 and 5 against light-quark or gluon jets, and $c$-quark jets, respectively. To compensate for differences between the $b$-tagging efficiencies and mis-tag rates in data and MC simulation, correction factors derived using $t\bar{t}$ events are applied to jets in the simulation as described in Refs.~\cite{btag1,btag2}.

Electron candidates used to veto events with prompt leptons in the hadron--hadron channel search are required to have $\pT > 10 \GeV$, $|\eta| < 2.47$ and to satisfy {\em loose} selection criteria on electromagnetic shower shape and track quality~\cite{Aad:2011mk}. Their longitudinal and transverse impact parameters must be within 2~mm and 1~mm of the primary vertex, respectively. 
In the lepton--hadron channel, further selections are applied. Electrons are required to satisfy the {\em tight} quality criteria, to have $\pT > 25 \GeV$, and to be isolated within the tracking volume. The electron identification efficiencies are of about 95\%, 91\% and 80\% for the {\em loose}, {\em medium} and {\em tight} working points respectively. The electron isolation requires that the scalar sum, $\Sigma \pT$, of the $\pT$ of inner detector tracks within a cone of size $\Delta R \equiv \sqrt{(\Delta\eta)^2+(\Delta\phi)^2} = 0.2$ around the electron candidate, is less than 10\% of the electron \pT. The tracks included in the scalar sum must have $\pT > 1 \GeV$, are matched to the primary vertex, and do not include the electron track.

Muon candidates are reconstructed using inner detector tracks either combined with muon spectrometer tracks or matched to muon segments~\cite{ATLAS-MuonRecoPerf2012}. They are required to have $\pT>10$~\GeV\  and $|\eta|<2.4$. Their longitudinal and transverse impact parameters must be within 1~mm and 0.2~mm of the primary vertex, respectively. 
These selections have an overall efficiency of about 99\%. Muon candidates that pass these selections are referred to as \emph{loose} muons and are used to veto events with prompt leptons in the hadron-hadron channel.
The candidates with $\pT > 25$~\GeV\ which fulfill the isolation requirement $\Sigma \pT < 1.8$~\GeV, i.e. with at most one additional track with $1<\pT < 1.8 $~\GeV\ reconstructed within a cone of size $\Delta R = 0.2$ around the muon track, 
are referred to as {\em tight} muons.

Event-level weights are applied to MC events to correct for differences between the lepton reconstruction and identification efficiencies measured in the simulation, and those measured in data.

Hadronically decaying tau lepton $(\tauhad)$ candidates are seeded by calorimeter jets with $\pT > 10$~\GeV. An $\eta$- and $\pT$-dependent 
energy scale calibration is applied to correct for the detector response and subtract energy from pile-up interactions~\cite{Aad:2014rga}. 
Tau lepton candidates are identified by using two boosted decision tree (BDT) algorithms that separate them from jets and electrons \cite{Aad:2014rga}. Variables describing the shower shape in the calorimeters and information from the tracking system are used to separate the collimated $\tauhad$ decay products from the generally broader jets resulting from quark and gluon hadronisation. Variables such as the number of tracks or the fraction of the total tau energy contained in a cone of size $\Delta R=0.1$ centred on the tau candidate provide strong discriminating power. To distinguish taus from electrons, the most discriminating characteristics are the transition radiation emitted by electrons in the TRT and the longer and wider shower generated by a hadronically decaying tau in the calorimeter compared with that produced by an electron. In addition to the two BDT selection criteria, a muon veto is applied. Hadronically decaying tau lepton candidates are required to have 
$\pT > 20$~\GeV, $| \eta | < 2.47$, and exactly one or three associated inner detector tracks (referred to as 1-prong and 3-prong candidates, respectively). The tau candidate is assigned an electric charge equal to the sum of the charges of the associated tracks, 
and this is required to be either +1 or -1. 
Three working points (\textit{loose}, \textit{medium}, and \textit{tight}) are used for each BDT. 
The hadron--hadron channel uses the tight identification working point for jet rejection and the medium identification working point for electron rejection, while the lepton--hadron channel uses the medium working point for both. The loose working point has been used to cross-check the background modelling. For the jet-veto BDT, the working points correspond to a signal efficiency of 70\%, 60\% and 40\% for 1-prong $\tauhad$, and 65\%, 55\% and 35\% for 3-prong $\tauhad$, respectively. The electron-veto BDT working points have a signal efficiency of 95\%, 85\% and 75\%, respectively. 
Efficiency scale factors are used to account for the mis-modelling of BDT input variables in the simulation. They are extracted by comparing efficiencies in data and simulation in a $Z\rightarrow\tau\tau$\ selection, using a tag-and-probe method described in Ref.~\cite{Aad:2014rga}.

As a given final-state particle can be simultaneously reconstructed as (for example) an electron, a jet and a hadronically decaying tau lepton, an algorithm is used to resolve such ambiguities. 
Electrons satisfying the medium quality criteria, 
muons satisfying the criteria described above except that on isolation, jets and hadronically
decaying tau candidates satisfying the selection criteria given above are considered by the algorithm. 
If two objects are close together in $\Delta R$, one of them is discarded according to the sequence specified in Table~\ref{tab:overlapremoval}. 
Electrons and muons close to jets, which are likely to originate from the decay of heavy-flavour hadrons, are removed from the list of leptons used in the analysis.  

\begin{table}
\begin{center}
\caption{Sequence of the overlap removal algorithm. 
Here, $\ell$ refers to electrons and muons. \label{tab:overlapremoval}}
\begin{tabular}{lc}
\hline
Condition & Discarded object \\     
\hline
$\DeltaR({\rm jet}, {\rm electron}) < 0.2$   & jet \\
$\DeltaR(\tauhad, \ell) < 0.2$  & $\tauhad$ \\
$\DeltaR({\rm jet}, \ell) < 0.4$  & $\ell$ \\
$\DeltaR(\tauhad, {\rm jet}) < 0.2$  & jet \\

\hline
\end{tabular}
\end{center}
\end{table}

The missing transverse momentum vector $\mathbf p^\mathrm{miss}_\mathrm{T}$, whose magnitude is referred to as  $E^\mathrm{miss}_\mathrm{T}$, is 
calculated as the negative vector sum of the transverse momenta of all reconstructed electrons, jets and muons, and calorimeter energy clusters not associated with any objects. For the $\mathbf p^\mathrm{miss}_\mathrm{T}$ computation, hadronically decaying taus are treated as jets. Clusters associated with electrons with $\pT> 10$~\GeV, and those associated with jets with $\pT >20$~\GeV\ are calibrated with the electron and jet cluster calibrations, respectively. For jets, the calibration includes the pile-up correction described earlier while the {\em jet vertex fraction} (JVF) requirement is not imposed. 
The JVF variable is the ratio of the sum of the transverse momentum of the tracks associated with the jet and originating from the selected primary vertex to the total $\pT$ sum of all tracks matched with the jet. This requirement rejects jets originating from pile-up. Clusters of energy deposits in calorimeter cells with $|\eta |< 2.5$ not associated with these objects are calibrated using both calorimeter and tracker information~\cite{Aad:2012re}.

\section{Event selection and background estimate}
\label{sec:selection}

\subsection{Hadron--hadron channel}
\label{sec:hadhadsel}

For the hadron--hadron channel search, events in the signal region are required to have exactly two oppositely charged hadronically decaying taus satisfying the tight identification criteria, no electrons or muons, and at least two jets with a 
JVF larger than 0.5 or $\pT > 50$~\GeV.  One of the jets 
must be $b$-tagged. The leading jet must also satisfy $\pT> 40$~\GeV. 

The missing transverse momentum must be larger than $150 \GeV$. The $\Delta \phi$ separation between each of the two leading jets and the direction of the missing transverse momentum must be greater than 0.5 radian, to suppress events where large \met\ arises from mis-measurement of jet energies.  
Beyond these preselection requirements, additional selections are made using transverse masses and derived variables, as explained below.
These selections have been determined using MC signal and background samples to maximise the expected significance of the signal.

The transverse mass associated with two final-state objects $a$ and $b$ is defined as
\begin{equation}
m_{{\rm T}} \left (\rm a, \rm b \right) = \sqrt{ m_{\rm a}^{2} + m_{\rm b}^{2} + 2 \left ( \ET^{\rm a} \ET^{\rm b} -  \mathbf{p_{\rm T}^a} . \mathbf{p_{\rm T}^b} \right )  }\,,
\label{transversemass}
\end{equation}
where $m$, $\ET$ and $\mathbf{p_{\rm T}}$ are the object mass, transverse energy and transverse momentum vector, respectively.
Objects entering the $m_{\rm T}$ calculation are always assumed to be massless, unless the transverse mass is used as part of a derived variable in the lepton--hadron channel (see Sect.~\ref{sec:selection-lephad}).

The {\em stransverse mass} ($m_{{\rm T}2}$)~\cite{Lester:1999tx,Barr:2003rg} is computed as
\begin{equation}
m_{{\rm T}2} \left ( \rm a, \rm b \right ) =
 \sqrt{\min_{\mathbf{ q^a_{\rm T} }+ \mathbf{ q^b_{\rm T} }= \mathbf{ p^{\rm{miss}}_{\rm T}} } \left ( \max \left [ 
    m^2_{\rm T} \left ( {\mathbf p^{\rm a}_{\rm T}}, {\mathbf q^{\rm a}_{\rm T}} \right ),  m^2_{\rm T} \left ( {\mathbf p^{\rm b}_{\rm T}},
    {\mathbf q^{\rm b}_{\rm T}} \right ) \right ] \right) } \,,
    \label{stransversemass}
\end{equation}
where $\mathbf q^{\rm a}_{\rm T}$ and $\mathbf q^{\rm b}_{\rm T}$ are vectors satisfying ${\mathbf q^{\rm a}_{\rm T} }+ {\mathbf q^{\rm b}_{\rm T} }= {\mathbf  p^{\rm{miss}}_{\rm T}}$, and the minimum is taken over all the possible choices of  $\mathbf q^{\rm a}_{\rm T}$ and $\mathbf q^{\rm b}_{\rm T}$.  

The selection criteria that define the signal region for the hadron--hadron channel (SRHH) rely on the following variables:

\begin{itemize}

\item $m_{{\rm T}2} (\tau_{{\rm had}},\tau_{{\rm had}})$ is defined using the momenta of the hadronically decaying taus and the missing transverse momentum, which is assumed to result from two invisible massless particles. The $m_{{\rm T}2} (\tau_{{\rm had}},\tau_{{\rm had}})$ variable is bounded from above by the $W$ boson mass for events where the two hadronically decaying taus originate from the decay of two W bosons and all the missing transverse momentum is carried by the neutrinos from the $W$ bosons decay, as is the case for the dominant background ($t\bar{t}$).

\item $m_{\rm T}^{\rm sum}(\tau_{{\rm had}},\tau_{{\rm had}})$ is defined as the sum of the transverse mass of each $\tauhad$\ candidate and the missing transverse momentum 
\begin{equation}
m_{\rm T}^{\rm sum} \left (\tau_{{\rm had}},\tau_{{\rm had}} \right ) = 
m_{\rm T} \left(\tau_{{\rm had1}} ,  \pT^{\rm{miss}}    \right)+m_{\rm T}\left(\tau_{{\rm had2}},  \pT^{\rm{miss}}   \right)
\end{equation}

The $m_{\rm T}^{\rm sum}(\tau_{{\rm had}},\tau_{{\rm had}})$ distribution is expected to reach higher values for the signal due to a larger number of invisible final-state particles than for the SM background processes.

\end{itemize}

For the SRHH signal region, the stransverse mass $m_{{\rm T}2} (\tau_{{\rm had1}},\tau_{{\rm had2}})$ is required to be larger than  50~\GeV\ while the $m_{\rm T}^{\rm sum}(\tau_{{\rm had1}},\tau_{{\rm had2}})$ variable is required to be larger than 160~\GeV. 
The signal selection efficiency, defined as the number of signal events that pass the full selection over the total number of generated events,
is only weakly dependent on the scalar tau mass, while it increases from 0.02\% to 0.7\% as the scalar top mass increases from 150~\GeV\ to 700~\GeV, for a \scalartau mass of 87~\GeV.  The distributions of $m_{{\rm T}2} (\tau_{{\rm had1}},\tau_{{\rm had2}})$ and $m_{\rm T}^{\rm sum}(\tau_{{\rm had1}},\tau_{{\rm had2}})$ are illustrated in Fig. \ref{fig:presel} after the preselection.

The background processes populating the SRHH selection are grouped into three categories. The first contains events with two real, hadronically decaying taus ({\em true} taus). It consists mainly of $t\bar{t}$ events, with smaller contributions from single-top-quark, $Z$+jets, diboson ($WW$, $WZ$, $ZZ$) and $t\bar{t}+V$ production, where $V=W,Z$. This set of backgrounds is estimated from simulation. The remaining backgrounds contain events where at least one tau candidate is an electron or a jet that passes the tau identification criteria ({\em fake} taus). The second category, which contains events with only one fake $\tauhad$, is composed of $t\bar{t}$, single-top-quark and $W$+jets events. The third and smaller category corresponds to processes with two fake taus. It is mostly composed of $t\bar{t}$, $Z(\rightarrow\nu\nu)$+jets, and single-top-quark events, which are all estimated from simulation. It has been verified that these backgrounds are well modelled: in kinematic selections where $t\bar{t}$ with true taus is expected to be the dominant process, the ratio of data over the MC prediction is compatible with one within systematics uncertainties.
The contribution from multi-jet events, where both tau candidates are fakes, is estimated from data using the jet smearing method described later in this section.

\begin{figure}[!hp]
\begin{center}
\includegraphics[width=0.80\textwidth]{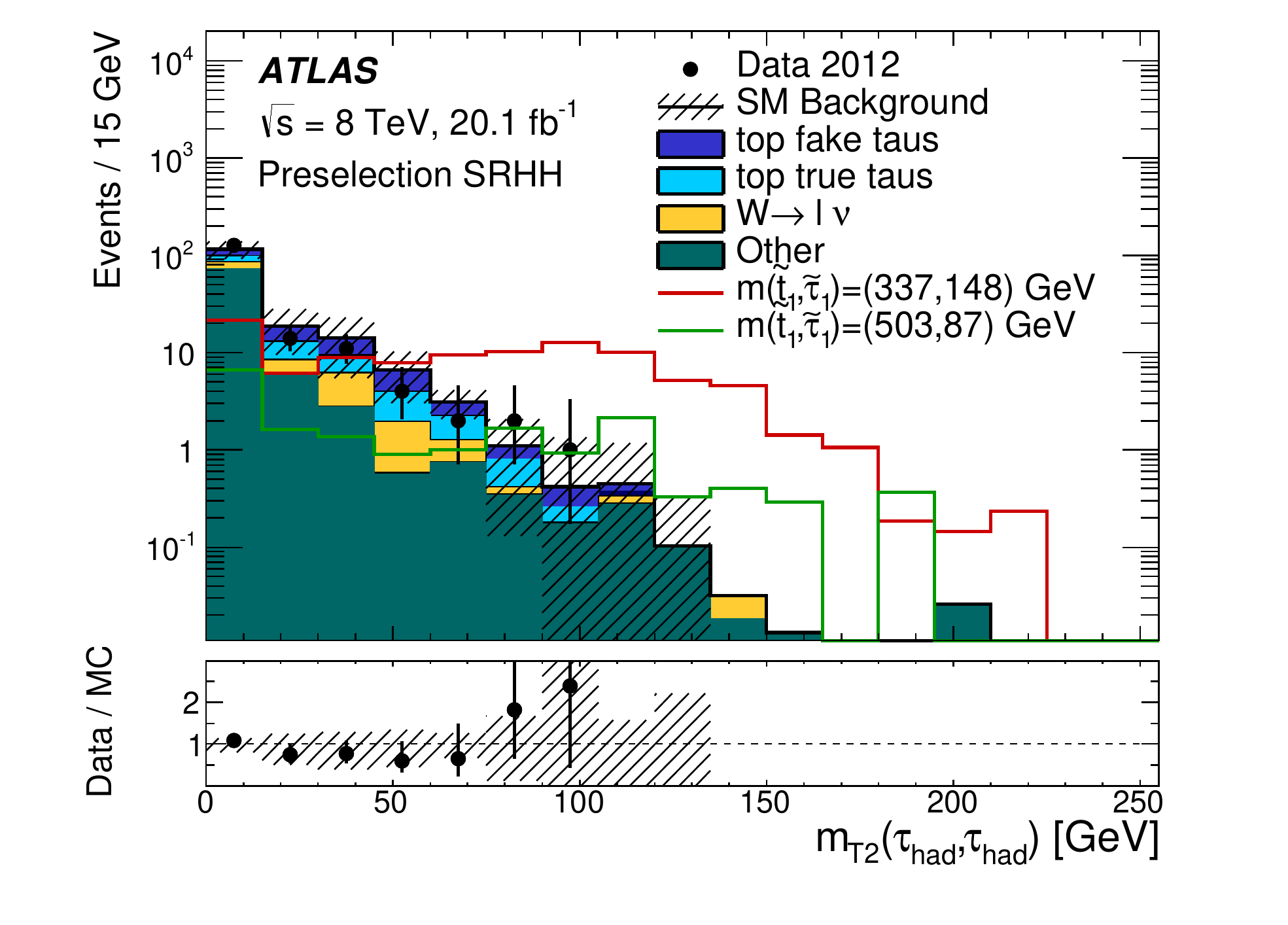}  \\
\includegraphics[width=0.80\textwidth]{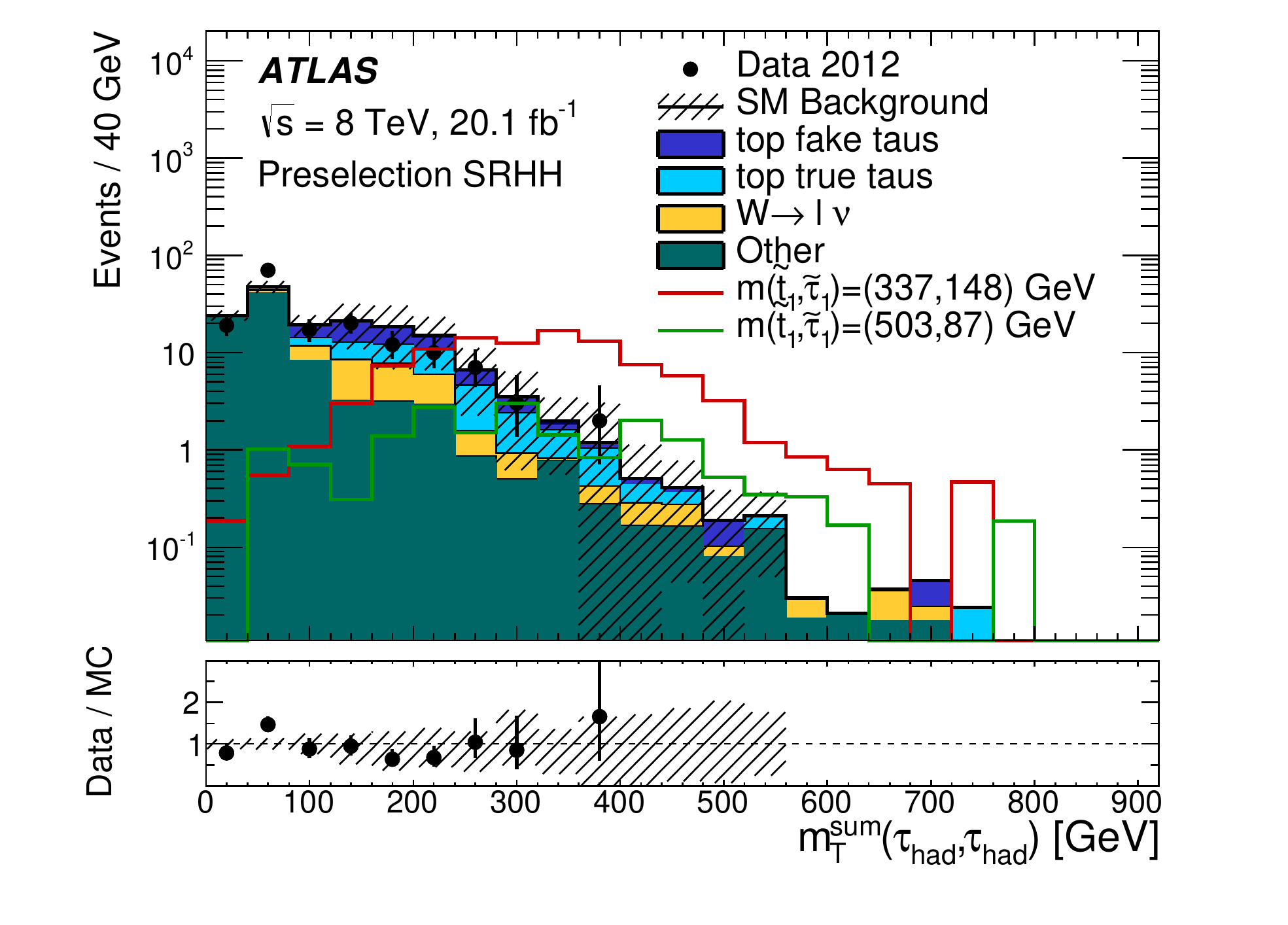}  
\end{center}

\caption{\label{fig:presel} Top: distribution of the stransverse mass constructed from the two $\tauhad$,  $m_{{\rm T}2}(\tau_{{\rm had}},\tau_{{\rm had}})$,  for events passing the hadron--hadron preselection requirements. Bottom: distribution of the sum of the transverse mass of each $\tauhad$ candidate and the missing transverse momentum, $m_{{\rm T}}^{\rm sum}(\tau_{{\rm had}},\tau_{{\rm had}})$, for events passing the hadron--hadron preselection requirements. The contributions from all SM backgrounds are shown as a histogram stack; the bands represent the total uncertainty. The distributions expected for two signal models are also shown.}

\end{figure}

The single-fake $\tauhad$\ backgrounds from top quark ($t\bar{t}$ and single-top) and $W$+jets events are estimated using MC simulations scaled to the observed number of data events in two dedicated control regions (CRHHTop and CRHHWjets).  These control regions require a single-muon trigger, one $\tauhad$\ satisfying the {\em tight} quality criteria, and one muon with $\pT>25$~\GeV\ that satisfies the {\em tight} quality criteria.  The $m_{{\rm T}2}$ and $m_{\rm T}^{\rm sum}$ variables are then calculated using the tau and muon momenta, considering the invisible particles as massless. One muon and one $\tauhad$\ are required in the control regions rather than two hadronically decaying taus in order to minimise signal contamination. Upper bounds are set on the $m_{{\rm T}2}$ and $m_{\rm T}^{\rm sum}$ variables, which make the contamination from the lepton--hadron signal negligible.  Table~\ref{tab:HHselections} details the selections defining the two control regions and the signal region.  The contributions to the background from the double-fake $\tauhad$\ sources are smaller than 4.5\% and therefore they are estimated using simulation without normalising to data in a control region.

A simultaneous likelihood fit is performed to determine the normalisation factors of the single-fake $\tauhad$ backgrounds, with the number of data events in each CR as constraint, and the systematic uncertainties described in Sect.~\ref{sec:systematics} included as nuisance parameters. The fit is used to predict the number of background events in the CRs and the SR. The background modelling is verified using two validation regions (VRs) by comparing the observed number of events in each VR with the number derived from the fit.  The single-fake $\tauhad$\ backgrounds from top quark and $W$+jets events each have a validation region, labelled VRHHTop and VRHHWjets. 
Like the control regions, they are defined using a muon and tau to avoid signal contamination, and the selections are summarised in Table~\ref{tab:HHselections}.  The validation regions are designed to be kinematically close to the signal region without overlapping with the control or signal regions.  The composition of the control and validation regions after the fit is shown in 
Fig.~\ref{fig:hadhad_pullHH}. The observed and expected background yields in the VRs are in good agreement, with 50 observed events in VRHHWjets ($48.5 \pm 6.9$\ expected) and 31 observed events in VRHHTop ($29.0 \pm 4.1$\ expected). 
It has also been verified that a normalisation factor for the top quark background with two real $\tauhad$ would be compatible with one within uncertainties.

The multi-jet background is estimated from data using the jet smearing method described in Ref.~\cite{Aad:2012fqa}.
A set of single-jet triggers is used to select a sample of events with at least two jets (of which at least one is required to be a $b$-jet), and two~$\tauhad$ candidates.
These events are required to have a low \met\ significance\footnote{The \met\ significance is defined as 
$ \met/\sqrt{\sum\limits_{\rm jets}E_{\rm T}+\sum\limits_{\rm soft\, terms}E_{\rm T}}$ 
 where soft terms correspond to clusters of energy deposits in the calorimeter which are not associated with any reconstructed object.}, 
to retain topologies where jets and tau candidates are well-balanced in the transverse plane and suppress processes with genuine \met.
The energy of jets and tau candidates is then smeared within the resolution of the calorimeter, in order to simulate \met\ arising from mis-measurements.
To minimise the statistical uncertainty, no identification criteria are applied to $\tauhad$\ candidates beyond the 1,3-track requirement, and a fake rate is used at a later stage to account for the tau identification efficiency.
The pseudo-dataset obtained after smearing serves as a template for the multi-jet background. Its normalisation is derived in a multi-jet-enriched CR, labelled CRHHQCD in Table~\ref{tab:HHselections}.
To estimate the background yield in the signal region, all SRHH requirements except the tau identification are applied to the normalised background template.
A weight is then applied to each event according to the probability for a jet reconstructed as a tau to satisfy the tight tau identification criteria. This fake rate is measured in data using events which fire a single-jet trigger, with at least two jets and a hadronically decaying tau candidate. It is found to be of the order of  1\% for 1-prong tau candidates and between 0.02\% and 0.4\% (with a strong $\pT$ dependence) for 3-prong tau candidates. The number of multi-jet events in the SR is estimated to be 
$ 0.0043 \pm 0.0007 \,({\rm stat}) \,^{+0.0039}_{-0.0008} \, ({\rm syst})$, and is therefore neglected.

\begin{table}[!h]
\begin{minipage}{\textwidth}
\caption{\label{tab:HHselections} Definition of the signal region (SRHH) for the hadron--hadron analysis. The selections of the associated control regions for $t\bar{t}$ and single-top-quark 
(CRHHTop) and $W$+jets (CRHHWjets) events with one fake hadronically decaying tau, as well as the validation regions (VRHHTop and VRHHWjets), are also shown. 
The $\ell$ entering the $m_{{\rm T}2}$ and $m_{\rm T}^{\rm sum}$ variables is either a $\tauhad$ (SR) or a muon (CRs and VRs).}
\begin{center}
\begin{small}

\begin{tabular}{ccccccccc}
\hline
Region & $N_{\tauhad}$ & $N_{\mu}$ & $N_{\rm jet}$ & $N_{b{\rm-jet}}$ & $E^\mathrm{miss}_\mathrm{T}$ &
 $\Delta\phi(j_{1,2},  \pT^{\rm{miss}}   )$ & $m_{{\rm T}2}(\tauhad,\ell)$ & $m_{\rm T}^{\rm sum}(\tauhad,\ell)$  \\
\hline
SRHH & 2 & 0 & $\geq$ 2 & $\geq$ 1 & $> 150$~\GeV & $\geq$ 0.5 & $> 50$~\GeV & $> 160$~\GeV\\
CRHHTop & 1 & 1 & $\geq$ 2 & $\geq$ 1 & $> 100$~\GeV & $\geq$ 0.5 & - & [70,120]~\GeV\\
CRHHWjets & 1 & 1 & $\geq$ 2 & 0 & $> 100$~\GeV & $\geq$ 0.5 & $< 40$~\GeV & [80,120]~\GeV\\
VRHHTop & 1 & 1 & $\geq$ 2 & $\geq$ 1  & $> 120$~\GeV & $\geq$ 0.5 & $< 40$~\GeV & [120,140]~\GeV\\
VRHHWjets & 1 & 1 & $\geq$ 2 & 0  & $> 120$~\GeV & $\geq$ 0.5 & $< 40$~\GeV & [120,150]~\GeV\\
CRHHQCD & $\geq$ 2\footnote{For the multi-jet control region (CRHHQCD), no identification criteria are applied to tau leptons.} & 0 & $\geq 2$ & $\geq1$ & $>150$~\GeV & $\leq0.5$\footnote{The $\Delta\phi$ requirement only applies to the sub-leading jet $j_2$.} & -& -\\
\hline

\end{tabular}
\end{small}
\end{center}
\end{minipage}
\end{table}

\begin{figure}[htb!]
\begin{center}
\includegraphics[width=0.8\textwidth]{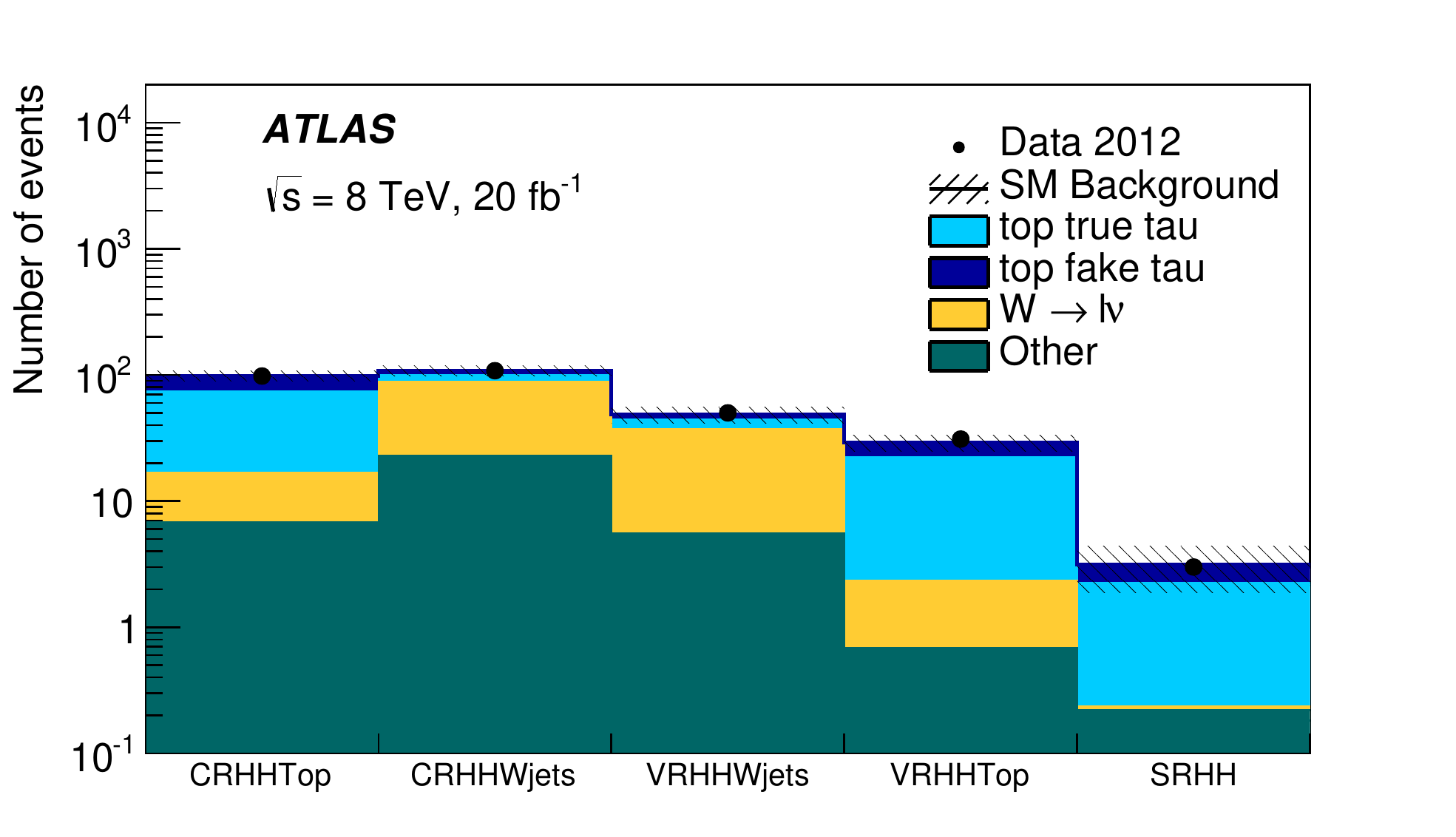}
\end{center}
\caption{Background yields and composition after the fit in the two CRs and the two VRs of the hadron--hadron channel analysis. Combined statistical and systematic uncertainties are shown as shaded bands. The observed number of events and the total (constrained) background are the same by construction in the CRs.} 
\label{fig:hadhad_pullHH}
\end{figure}

\subsection{Lepton--hadron channel}
\label{sec:selection-lephad}

The search in the lepton--hadron channel requires exactly one hadronically decaying tau, exactly one isolated electron or muon with $\pT > 25$~\GeV, and no 
further isolated electrons or muons with $\pT > 10$~\GeV.  The hadronically decaying tau and the lepton are required to have opposite electric charge. 
Each event must also contain at least two jets, where at least one of the two jets must have $\pT > 50 $~\GeV, and at least one of the two must be $b$-tagged.  

After this common preselection, two different signal regions are defined to target signal models with a \scalartop mass large or small in comparison to the top-quark mass. These are referred to as the {\em low-mass} (SRLM) and {\em high-mass} (SRHM) selections in the following, and they have been optimised with respect to the expected significance of the signal.
The selections for the two signal regions are summarised in Tables~\ref{tab:LHLM} and~\ref{tab:LHHM}.  The low-mass selection requires a second $b$-jet. Three \mttwo variables are employed in the selections, with different choices of the two visible four-momenta used in the calculation from Eq.~(\ref{stransversemass}):

\begin{itemize}

\item  $\mttwo(\ell, \tauhad)$ uses the momenta of the light lepton and the hadronically decaying tau. The missing transverse momentum is assumed to result from two invisible massless particles. The $\mttwo(\ell, \tauhad)$ variable is bounded from above by the $W$ boson mass for events where the light lepton, the hadronically decaying tau and the missing transverse momentum originate from the decay of a pair of $W$ bosons, which is the case for most of the background ($\ttbar$\ and $Wt$).  The high-mass selection requires this variable to be large, because its distribution for signal models with heavy scalar taus and scalar tops peaks at higher values than for the top-quark-dominated SM background.

\item $\mttwo(b\ell, b\tauhad)$ is calculated using the two jets with the highest $b$-tagging weight. 
One of them is paired with the light lepton and the other with the $\tauhad$.  The four-momentum vectors of 
the two resulting particle pairs are then used in the \mttwo algorithm.  The missing transverse momentum is assumed to be carried 
by two invisible massless particles. For $t\bar t$ events where the 
jet and the lepton belong to the decay of the same top quark, this variable is 
bounded from above by the top-quark mass. Similarly, for signal events, the upper bound on this variable is the \scalartop mass. 
A maximum-value cut is therefore used in the low-mass 
selection and a minimum-value cut in the high-mass selection. The calculation of the variable requires
the resolution of a two-fold ambiguity in the pairing of the jets and the leptons. Only the pairings for which 
$m(b\ell)$ and $m(b\tauhad)$ are both smaller than $m_t$ are considered.\footnote{For top-quark pair production events 
where the lepton and the jet belong to the decay of the same top quark, the invariant mass has an 
upper bound at $\sqrt{m_t^2-m_W^2}$, approximately 152~\GeV. The algorithm tries to select pairs that satisfy this condition, loosened to 
account for the detector resolution.} 
If exactly one pairing satisfies the condition, that pairing is used in the $\mttwo$ calculation. 
If both pairings satisfy the condition, $\mttwo$ is calculated 
for both pairings and the smaller value is taken. If no pairing satisfies the condition,  
the event is considered to have passed the  $\mttwo(b\ell, b\tauhad)$ 
selection for the high-mass signal region and to have failed it for the low-mass signal region.

\item $\mttwo(b\ell, b)$ is only used for the low-mass selection.
 The system of one of the $b$-jets and the light lepton is considered 
as the first visible four-momentum. Only pairings for which $m(b\ell) < m_t$ are considered.
If neither pairing satisfies the condition, the event is discarded, while if both pairings do, the 
pairing which yields the smaller value of $ \mttwo(b\ell, b)$ is used. The invisible particle associated 
with this system is assumed to be massless. The other $b$-jet is the 
second visible system used in the $\mttwo$ calculation, and the mass of the associated 
invisible particle is set to the $W$ boson mass, as the algorithm targets $t \bar t$ events where
one lepton from a $W$ boson decay is not detected or identified.  For the dominant top-quark background,
the $\mttwo(b\ell, b)$\ variable is bounded from above by the top-quark mass.  This variable has a softer 
distribution for low-mass signal events than the background, and a maximum-value cut of 100~\GeV\
is applied.

\end{itemize}

\begin{figure}[hp!]
\begin{center}
\includegraphics[width=0.80\textwidth]{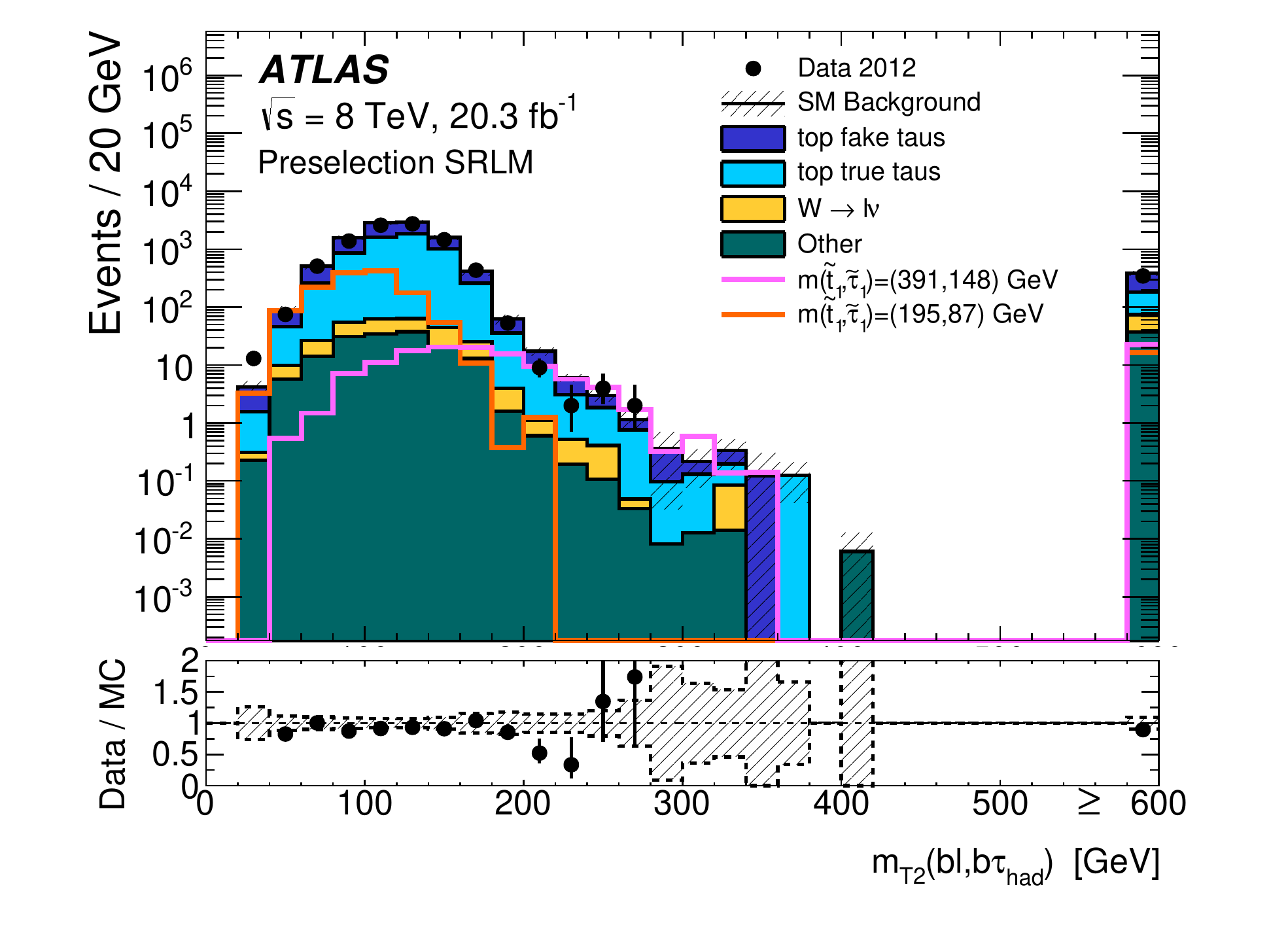} \\
\includegraphics[width=0.80\textwidth]{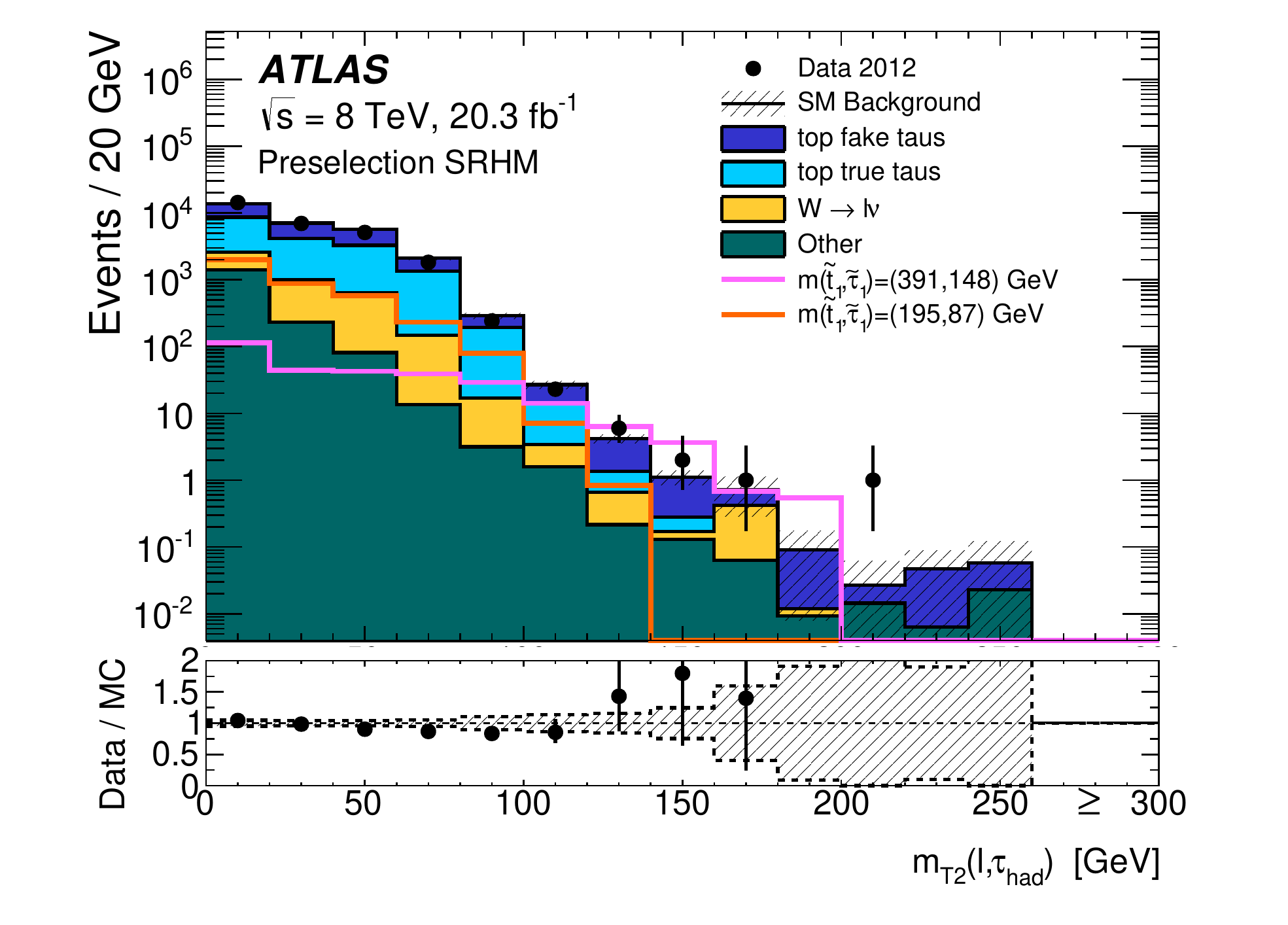}
\end{center}
\caption{Top: distribution of the stransverse mass constructed from the $b$-jet plus lepton and $b$-jet plus $\tauhad$, $m_{{\rm T}2}(b \ell, b \tauhad)$, for events passing the lepton--hadron preselection requirements with the additional requirement of a second $b$-tagged jet.  Bottom: distribution of the stransverse mass constructed from the momenta of the light lepton and the hadronically decaying tau, $m_{{\rm T}2}(\ell, \tauhad)$, for events passing the lepton--hadron preselection requirements. The contributions from all SM backgrounds are shown as a histogram stack; the bands represent the total uncertainty. 
The overflow bin in the $m_{{\rm T}2}(b \ell, b \tauhad)$ plot is filled with the events that have no $(b \ell, b \tauhad)$ pairing satisfying $m(b\ell) < m_t$ and $m(b\tauhad) <m_t$.
The distributions expected for two signal models are also shown.} 
\label{fig:lephadPresel}
\end{figure}

The distributions for $m_{{\rm T}2}(b \ell, b \tauhad)$\ and $m_{{\rm T}2}(\ell, \tauhad)$ are illustrated in Fig.~\ref{fig:lephadPresel}  after the preselection, showing the separation between two signal models and the SM background.   The $m_{{\rm T}2}(b \ell, b \tauhad)$\ variable is used to distinguish the \scalartop signal from the dominant top-quark backgrounds for both the low-mass and high-mass selections.

Another variable used in the selections is the ratio of the scalar sum of the transverse momenta of 
the two leading jets ($H_{\rm T}$) to the effective mass, $m_{\rm eff} = E_{\rm T}^{\rm miss } + H_{\rm T} + p_{\rm T}^{\ell} + p_{\rm T}^{\tauhad}$, where $p_{\rm T}^{\ell}$ and $p_{\rm T}^{\tauhad}$ are the transverse momenta of the lepton and the hadronically decaying tau, respectively. This ratio, $H_{\rm T}/m_{\rm eff}$, tends to be smaller for signal events 
because of the high number of invisible particles in the final state, and it is required to be less than 0.5. 
The high-mass selection also requires the missing transverse momentum to be larger than 150~\GeV\ and $m_{\rm eff}$ to be larger than 400~\GeV\ because the decay products of a high-mass \scalartop\ would have large momenta.  The low-mass selection requires $(p_{\rm T}^{\ell} + p_{\rm T}^{\tauhad})/m_{\rm eff} > 0.2$\ because the difference between the masses of the \scalartop\ and \scalartau\ is relatively small in comparison to the difference between the masses of the top quark and the $W$\ boson.  Finally, the $m_{\rm T}(\ell, \pT^{\rm{miss}}  )$ variable is used to distinguish events with real tau leptons from events with fake tau leptons in the dominant top-quark background, and to distinguish multi-jet events from $W$+jets events.
The definitions of the low-mass and high-mass SRs are summarised in Tables~\ref{tab:LHLM} and~\ref{tab:LHHM}, respectively.

The signal selection efficiency of the low-mass selection 
is between 0.008\% and 0.01\% for the models with a scalar top mass between 150~\GeV\ and 200~\GeV, which is the target of this selection. The signal efficiency of the high-mass selection increases with the scalar top mass. For a fixed scalar top mass, it increases with the scalar tau mass as the  $m_{{\rm T}2}(\ell, \tauhad)$ selection becomes more efficient, up to the region with $m(\tilde t_1) - m(\tilde \tau) < 50$~\GeV\ where the $b$-jets become too soft to be efficiently detected. Outside this region, which is better targeted by the lepton--lepton channel, the efficiency of the high-mass selection varies between 0.0007\% and 1\% for a scalar top mass between 200~\GeV\ and 700~\GeV. 

\begin{table}[!h]
\caption{\label{tab:LHLM} Definition of the signal region SRLM used in the low-mass
lepton--hadron analysis. The selections of the associated control regions for top-quark events with true taus 
(CRTtLM), top-quark events with fake taus (CRTfLM), and $W$+jets (CRWLM), and of the validation region 
(VRTLM) are also given.}
\begin{center}
\begin{tabular}{cccccccc}
\hline
\rule{0pt}{3.5ex}   Region   & $N_{b{\rm -jet}}$ & $H_{\rm T}/m_{\rm eff}$ & $\frac{p_{\rm T}^{\ell}+p_{\rm T}^{\tauhad}}{m_{\rm eff}}$ & $m_{{\rm T}2}(b \ell, b)$ & $m_{{\rm T}2}(b \ell, b \tauhad)$ &  $m_{\rm T}(\ell,\pT^{\rm{miss}})$ & $m_{\rm eff}$\\[0.5ex]
\hline
SRLM         & $\ge 2$ & $< 0.5$ & $> 0.2$ & $< 100$~\GeV\ &  $< 60$~\GeV\ &- & - \\
CRTtLM     & $\ge 2$ & -              & $> 0.2$ & $< 100$~\GeV\ & $110-160$~\GeV\ & $> 100$~\GeV\  & - \\
CRTfLM     & $\ge 2$ & -             & $> 0.2$ & $< 100$~\GeV\ & $110-160$~\GeV\ & $< 100$~\GeV\ & - \\
CRWLM     &      0       & $<0.5$ & $> 0.2$ & -                  & -     & $> 40$~\GeV\ & $< 400$~\GeV\ \\
VRTLM       & $\ge 2$ & $> 0.5$ & $> 0.2$ & $< 100$~\GeV\ &  $60-110$~\GeV\ & - & - \\
\hline
\end{tabular}
\end{center}
\end{table}

\begin{table}[!h]
\caption{\label{tab:LHHM} Definition of the signal region SRHM used in the high-mass
lepton--hadron analysis. The selections of the associated control regions for top-quark events with true taus 
(CRTtHM), top-quark events with fake taus (CRTfHM), and $W$+jets (CRWHM), and of the validation region 
(VRTHM) are also given.}
\begin{center}
\begin{tabular}{cccccccc}
\hline
Region   & $N_{b{\rm -jet}}$ & \MET & $m_{\rm eff}$ & $H_{\rm T}/m_{\rm eff}$ & $m_{{\rm T}2}(b \ell, b \tauhad)$ & $m_{{\rm T}2}(\ell, \tauhad)$ &  $m_{\rm T}(\ell,\pT^{\rm{miss}})$ \\
\hline
SRHM         & $\ge 1$ & $> 150$~\GeV\ & $> 400$~\GeV\ & $< 0.5$ & $> 180$~\GeV\ &  $> 120$~\GeV\ & - \\
CRTtHM     & $\ge 1$ & $> 150$~\GeV\ &  $> 400$~\GeV\ & $< 0.5$ & $> 180$~\GeV\ & $20$-$80$~\GeV\ & $> 120$~\GeV \\
CRTfHM     & $\ge 1$ & $> 150$~\GeV\ &  $> 400$~\GeV\ & $< 0.5$ & $> 180$~\GeV\ & $20$-$80$~\GeV\ & $< 120$~\GeV \\
CRWHM     &      0       & $> 150$~\GeV\ &  $> 400$~\GeV\ & $< 0.5$ & -                         & $20$-$80$~\GeV\ & $40$-$100$~\GeV \\
VRHM         & $\ge 1$ & $< 150$~\GeV\ & $> 400$~\GeV\ & $< 0.5$ & $> 180$~\GeV\ &  $> 80$~\GeV\ & - \\
\hline
\end{tabular}
\end{center}
\end{table}

In the lepton--hadron channel, the ratio of real to fake hadronically decaying tau events depends on the 
background process. In $W$+jets events, the light lepton is always a real lepton from 
the $W$ decay, due to the high reconstruction efficiency and purity of final-state electrons and muons,
while the $\tauhad$\ is faked by a recoiling hadronic object. In $\ttbar$ and $Wt$ events, the 
light lepton originates from the decay of one of the $W$ bosons while the hadronically decaying 
tau candidate can be either a real or a fake tau. These processes ($W$+jets, $\ttbar$, and $Wt$) are 
the main background sources and are estimated by MC simulation scaled to the observed data 
in three CRs for each SR.  The CRs are enriched in either $W$+jets, top-quark events with true hadronically decaying taus, 
or top-quark events with fake hadronically decaying taus (where the top-quark events include both single and pair production), and are used to derive normalisation factors for these three 
categories of background. For the low-mass selection SRLM, the true- and fake-tau top-quark backgrounds are controlled by CRTtLM and CRTfLM, while CRWLM controls the $W$+jets background. For the high-mass selection SRHM, the three control regions CRTtHM, CRTfHM and CRWHM are used to normalise the true- and fake-tau top-quark backgrounds and the $W$+jets background.
The CRs are defined in Table~\ref{tab:LHLM} for the low-mass selection 
and in Table~\ref{tab:LHHM} for the high-mass selection. The minor contribution from 
other background processes is estimated from simulation. 

A simultaneous likelihood fit is performed to obtain the three normalisation factors for each SR, using the observed number of data events in each CR as constraints, and with the systematic uncertainty
sources (described in Sect.~\ref{sec:systematics}) treated as nuisance parameters. The fit is used to predict the
number of background events in the CRs and the SR. The validity of the 
background modelling is verified by using a validation region for each SR
and comparing the observed number of events with the prediction from the
fit. For the low-mass selection, the validation region VRLM is defined in Table~\ref{tab:LHLM}, 
while the validation region VRHM is defined in Table~\ref{tab:LHHM} for the high-mass selection. 

The background composition and the observed number of events in each CR as well as in the VR and SR are shown in Fig.~\ref{fig:lephad_pullLM} for the low-mass selection and in Fig.~\ref{fig:lephad_pullHM} for the high-mass selection. The observed and expected background yields in the VRs are in good agreement, with 386 observed events for the low-mass selection ($351 \pm 84$\ expected) and 17 observed events in the high-mass selection ($22 \pm 5$\ expected). The expected background yields and observed number of events in the SRs are reported in Sect.~\ref{sec:results}.

\begin{figure}[htb!]
\begin{center}
\includegraphics[width=0.8\textwidth]{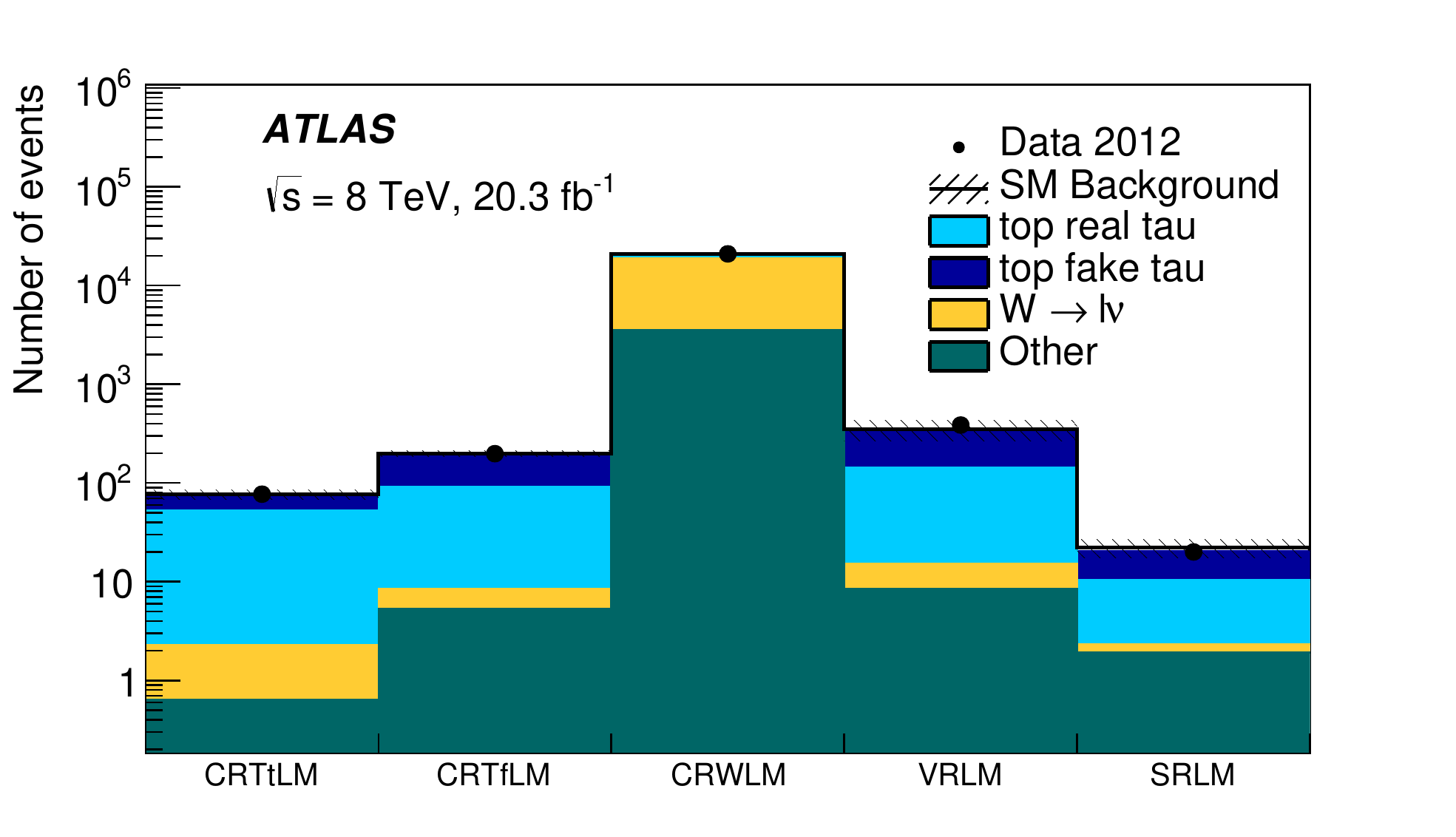}
\end{center}
\caption{Background yields and composition after the fit for the three CRs and the VR in the lepton--hadron channel low-mass selection. Combined statistical and systematic uncertainties are shown as shaded bands. The observed number of events and the total (constrained) background are the same by construction in the CRs.} 
\label{fig:lephad_pullLM}
\end{figure}

\begin{figure}[htb!]
\begin{center}
\includegraphics[width=0.8\textwidth]{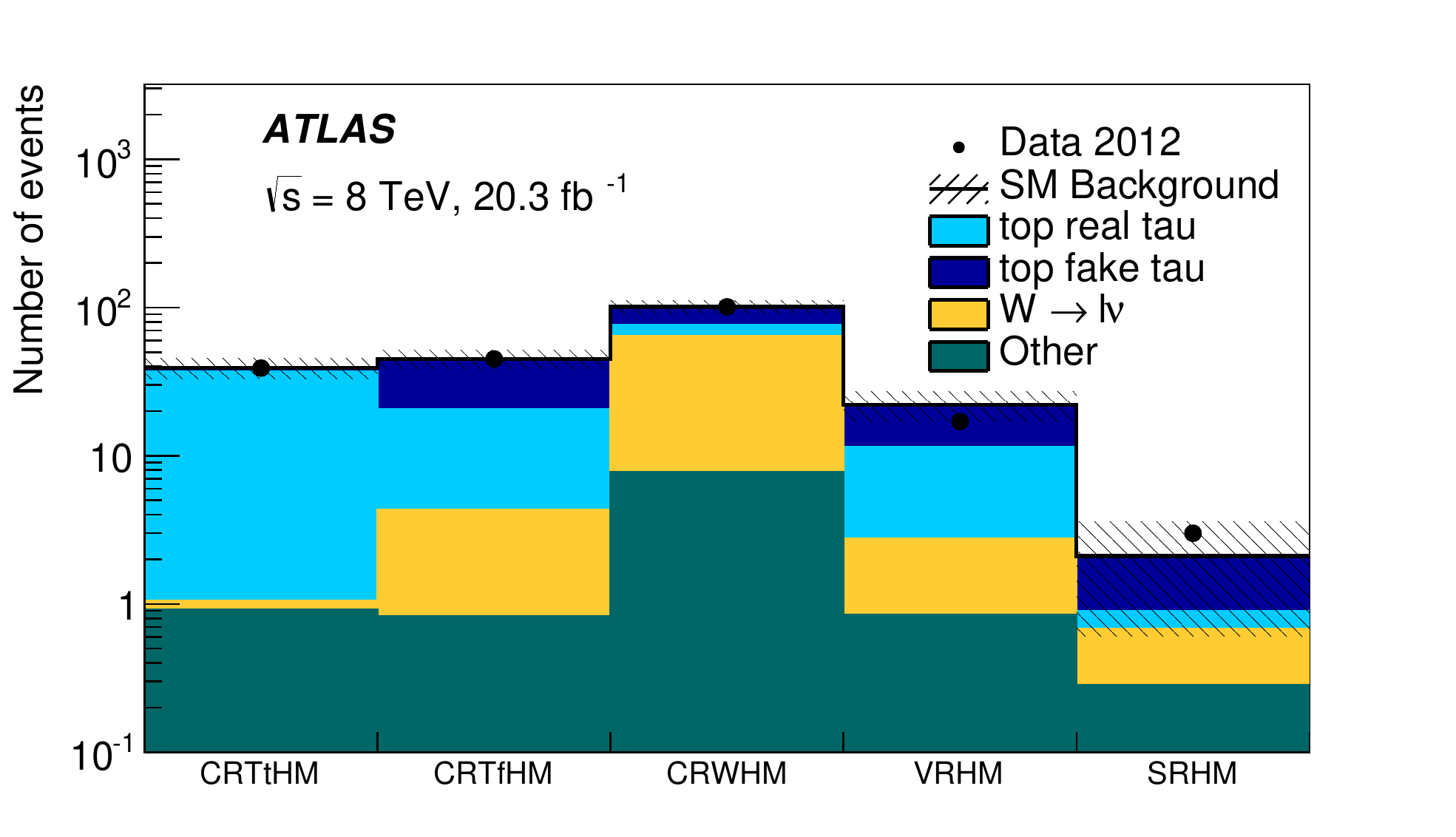}
\end{center}
\caption{Background yields and composition after the fit for the three CRs and the VR in the lepton--hadron channel high-mass selection. Combined statistical and systematic uncertainties are shown as shaded bands. The observed number of events and the total (constrained) background are the same by construction in the CRs.} 
\label{fig:lephad_pullHM}
\end{figure}

The background estimate with fake hadronically decaying taus (either from top-quark or $W$+jets events) is validated using an alternative method.
The observed rate of events with a light lepton and a $\tauhad$\ with the same electric charge is scaled by the expected ratio of  
opposite-sign (OS) to same-sign (SS) events for the fake $\tauhad$\ backgrounds, which is estimated from 
MC simulation. 
Too few SS events are observed for the SRHM selection to make a meaningful prediction, so the method is only viable for the looser SRLM 
selection, for which it predicts $12 \pm 6$ events with fake hadronically decaying taus, in 
agreement within uncertainties with the sum of $W$+jets and top-quark events with fake hadronically decaying taus obtained from the fit, 
which is $12 \pm 5$ events. 

\section{Systematic uncertainties}
\label{sec:systematics}

Various sources of systematic uncertainty affecting the predicted background yields in the signal regions are considered. 
The uncertainties are either computed directly in the SR when backgrounds are estimated from simulation, or propagated through the fit for backgrounds that are normalised in CRs.

The dominant detector-related systematic uncertainties considered in these analyses are the jet energy scale and resolution~\cite{Aad:2011he}, the $\tauhad$ 
energy scale and BDT identification efficiency~\cite{Aad:2014rga}, and the $b$-tagging efficiency~\cite{btag1,btag2}. The energy scale and resolution of clusters in the calorimeter not associated with electrons, muons or jets, which affect the missing transverse momentum calculation, are also a source of systematic uncertainty. In all cases, the difference
in the predicted background or signal between the nominal MC simulation and that obtained after applying each systematic variation is used to determine the 
systematic uncertainty on the background or signal estimate. Parts of the systematic uncertainties cancel when a background is estimated from a control region, 
but they do not cancel for processes normalised to their theoretical cross section.
The remaining detector-related systematic uncertainties, such as those on lepton reconstruction efficiency and on the modelling of the trigger, are of the order of a few percent. A 2.8\% uncertainty on the luminosity determination was measured using techniques similar to that of Ref.~\cite{lumi},
and it is included for the normalisation of all signal and background MC samples. The signal uncertainties are between 10\% and 15\% for models close to the observed exclusion contour. 

Various theoretical uncertainties are considered for the modelling of the major SM backgrounds. 
In the case of top-quark contributions, the predictions of {\tt POWHEG-BOX} are compared with those of {\tt MC@NLO-4.06}  to estimate the uncertainty due to the choice of generator.  
The difference in the yields obtained from {\tt POWHEG-BOX} interfaced to {\tt PYTHIA} and {\tt POWHEG-BOX} interfaced to {\tt HERWIG} is taken as the systematic uncertainty due to parton shower modelling, and the predictions of dedicated {\tt ACERMC-3.8} samples generated with different tuning parameters are compared to give the uncertainty related to the modelling of initial- and final-state radiation (ISR/FSR).
At NLO, contributions with an additional bottom quark in the final state lead to ambiguities in the distinction between the $Wt$ process ($gb\rightarrow Wt b$) and top-quark pair production. All the $Wt$ samples, generated using {\tt MC@NLO-4.06} and {\tt POWHEG-BOX}, use the diagram removal scheme~\cite{Frixione:2008yi} to model this interference. The {\tt ACERMC-3.8} event generator is used to simulate the $WWb$ and $WWb\bar{b}$ final states at leading order (which include both the $t\bar{t}$ and $Wt$ single-top-quark processes); the predictions of these {\tt ACERMC-3.8} samples are then compared to those of the nominal MC samples 
in order to assess the uncertainty on the background estimate from this interference.
The uncertainties on $W$+jets and $Z$+jets production are evaluated by studying the predictions of {\tt ALPGEN-2.14} with various choices of the renormalisation and factorisation scales.

The impact of systematic uncertainties on the total background estimate in the different SRs is shown in Table~\ref{tab:uncertainties-cutandcount}. The table quotes, for each SR, the relative background uncertainty attributed to each source. 

 \begin{table} [htb!]
\caption{\label{tab:uncertainties-cutandcount} Summary of background estimates and the associated total uncertainties.The size of each systematic uncertainty is quoted as a relative uncertainty on the total background. A dash indicates a negligible contribution to the uncertainty. The individual uncertainties can be correlated, and thus do not necessarily sum in quadrature to the total 
relative uncertainty.}

\begin{center}
\begin{tabular}{lccc}
\hline
                                       &  SRHH &  SRLM & SRHM \\
                                        \hline
Background events & $3.1 \pm 1.2$ & $22.1 \pm 4.7$ & $2.1 \pm 1.5$ \\
\hline
\multicolumn{4}{l}{Uncertainty Breakdown [$\%$]:}\\
\hline
Jet energy scale and resolution            & 17 &  13 & 2 \\
Tau energy scale                                   & 9 &   4 & 3 \\
Cluster energy scale and resolution      & 1 &   2 & 4 \\
$b$-tagging				            & 2 &   4  & 2 \\

Top-quark theory uncertainty                & 37 & 11 & 64   \\
$W$+jets theory and normalisation      & -   &  1 & 19 \\

Simulation statistics                              & 20 & 6 & 21    \\
Top normalisation                                & 18  & 6 & 20     \\

\hline
\end{tabular}
\end{center}
\end{table}

Signal cross sections are calculated at NLO+NLL with a total associated uncertainty between 14\% and 16\% for scalar top masses between 150~\GeV\ and 560~\GeV.

\section{Results and interpretation}
\label{sec:results}

\begin{table}[htb!]
\caption{\label{tab:SR} Observed number of events and background fit results for the hadron--hadron SR and the two lepton--hadron SRs.  Combined statistical and systematic uncertainties are given. The uncertainties between the different background components can be correlated, so they do not necessarily sum to the total background uncertainty.  A dash indicates a negligible background contribution. The nominal expectations from MC simulation are given for comparison in the lower part of the table.}
\begin{center}
{\small
\begin{tabular}{lrrr}
\noalign{\smallskip}\hline\noalign{\smallskip}
{\bf  Channel}                                      & SRHH     & SRLM       & SRHM             \\[-0.05cm]
\noalign{\smallskip}\hline\noalign{\smallskip}
Observed events                                   &    3      &          20           &   3    \\
\noalign{\smallskip}\hline\noalign{\smallskip}
Total (constrained) background events          & $3.1 \pm 1.2$   & $ 22.1 \pm 4.7$ & $ 2.1 \pm 1.5$   \\
\noalign{\smallskip}\hline\noalign{\smallskip}
Top with only true tau(s)                 & $2.0 \pm 1.1$       & $ 8.2 \pm 3.9$          & $0.2^{+0.3}_{-0.2}$     \\[0.8ex]
Top with at least one fake tau          & $0.9 \pm 0.5 $      & $9.8 \pm 4.5$          & $1.2^{+1.4}_{-1.2}$   \\[0.8ex]
$W$+jets                                       &   $0.01_{-0.01}^{+0.02}$                      & $2.2 \pm 0.6$          & $0.4 \pm 0.4$       \\[0.8ex]
$Z/ \gamma^{*}$+jets&       0.04 $^{+0.15}_{-0.04} $                  & $1.9 \pm 1.1$          &          --                  \\[0.8ex]
$\ttbar +V$                                      &   0.04 $\pm 0.02$                       &          --                   &   $0.3 \pm 0.1$               \\[0.8ex]
Diboson                                        &   $0.14\pm0.02$  &           --                   &         --               \\[0.8ex]
\noalign{\smallskip}\hline\noalign{\smallskip}
	Expected background events before the fit     & 3.7        &     25.8  &    2.2            \\[0.8ex]
\noalign{\smallskip}\hline\noalign{\smallskip}
	Top with only true tau(s)           &    2.0    & 11.5    &  0.18 \\[0.8ex]
	Top with at least one fake tau      &     1.4   & 10.1   &  1.1  \\[0.8ex]
	$W$+jets                            &       0.01    & 2.4      &   0.65\\[0.8ex]
	$Z/ \gamma^{*}$+jets                & 0.04         & 1.9     &    --    \\[0.8ex]
	$\ttbar + V$                           &        0.04  &      --           &  0.27          \\[0.8ex]
	Diboson                             &   0.14    &      --           &    --      \\[0.8ex]

\noalign{\smallskip}\hline\noalign{\smallskip}
\end{tabular}
}
\end{center}
\end{table}

\begin{table}[htb!]
\caption{\label{tab:upperlimits} Left to right: Total constrained background yields, number of observed events, 95\% CL observed (expected) upper limits on the number of BSM events, $S_{\rm obs. (exp.)}^{95}$, and the visible cross section, $\langle\mathcal{A}\epsilon\sigma\rangle_{\rm obs. (exp.)}^{95}$.
}
\begin{center}
\begin{tabular}{ccccc}

\hline \\[-2ex]   
{Signal Region}  & Background & Observation & $S_{\rm obs. (exp.)}^{95}$  &  $\langle\mathcal{A}\epsilon\sigma\rangle_{\rm obs. (exp.)}^{95}$ [fb] \\ [0.8ex]
\hline \\[-2ex]   
SRHH      & $3.1 \pm 1.2$    & 3   &   5.5 $\left(5.5^{+2.1}_{-1.3}\right)$ & 0.27 $\left(0.27^{+0.11}_{-0.06}\right)$  \\[0.8ex]
SRLM & $22.1 \pm 4.7$  & 20 & 12.4 $\left(13.2^{+4.9}_{-3.5}\right)$ & 0.61 $\left(0.65^{+0.24}_{-0.17}\right)$   \\[0.8ex]
SRHM & $2.1 \pm 1.5 $  & 3   &   6.4 $\left(5.2^{+2.6}_{-0.9}\right)$ & 0.31 $\left(0.26^{+0.13}_{-0.04}\right)$   \\ [0.8ex] 
\hline

\end{tabular}

\end{center}
\end{table}

\begin{table}
\caption{\label{tab:acceptanceEfficiency} Acceptance times efficiency for the various signal regions, for a few selected (\scalartop, \scalartau) signal mass hypotheses. For each mass point, values are shown only for channels targeting that point. The lepton--lepton results are taken from Ref.~\cite{Aad:2014qaa}.}
\begin{center}
\begin{tabular}{cccccc}
\hline
$\stopOne$\ mass & $\stauOne$\ mass & lepton--lepton &  lepton--hadron  &  lepton--hadron & hadron--hadron \\
$[{\rm GeV}]$ & $[{\rm GeV}]$ & $A \times \epsilon$  &  $A \times \epsilon$ (SRLM)  & $A \times \epsilon$ (SRHM) & $A \times \epsilon$ \\ [0.8ex]
\hline
153  &  87    & -   & $1.29\times10^{-4}$  &  -  &   $2.27\times10^{-4}$ \\
195  &  87    & -   & $1.36\times10^{-4}$  & -  &   $4.46 \times10^{-4}$  \\
195  &  148   & $1.71 \times 10^{-4}$  & $7.80\times10^{-5}$  & - & $7.00 \times 10^{-4}$ \\
195  &  185   & $8.01 \times 10^{-4}$  &         -            & -     &    -   \\
391  &  148   & $7.32 \times 10^{-4}$   & -  & $9.44\times10^{-4}$  & $3.40 \times 10^{-3}$ \\
503  &  493   & $1.03 \times 10^{-2}$  &         -            & -        &         -  \\
561  &  87    & -  & -  & $1.74\times10^{-3}$  &    $6.70 \times 10^{-3}$ \\
561  &  337   & -   & -  & $1.30\times10^{-2}$ &   $9.90 \times 10^{-3}$  \\
561  &  500   & -  & -  & $8.68\times10^{-3}$ &   $2.50 \times 10^{-3}$  \\
\hline
\end{tabular}
\end{center}
\end{table}

The numbers of events observed in the hadron--hadron SR and in the two lepton--hadron SRs are reported in Table~\ref{tab:SR}, 
along with the background yields  before and after the background-only likelihood fit. 
In both the results and interpretation tables (Tables~\ref{tab:SR}\ and \ref{tab:upperlimits}) the quoted uncertainties include all the sources of statistical and systematic uncertainty. Good 
agreement is seen between the observed yields and the background estimates.

Figure~\ref{fig:SR} shows the distributions of $m_{\rm T}^{\rm sum}(\tauhad,\tauhad)$ and $m_{{\rm T}2}(\tauhad,\tauhad)$\ for the hadron--hadron channel, for events 
satisfying all the SR criteria except that on the variable being reported in the figure. Figure~\ref{fig:SR2} shows $m_{{\rm T}2}(b \ell, b \tauhad)$ for the lepton--hadron
low-mass selection and $m_{{\rm T}2}(\ell, \tauhad)$ for the lepton--hadron high-mass selection for events satisfying all
the corresponding SR criteria except those on the variable displayed in the figure.

\begin{figure}[hp!]
\begin{center}
\includegraphics[width=0.80\textwidth]{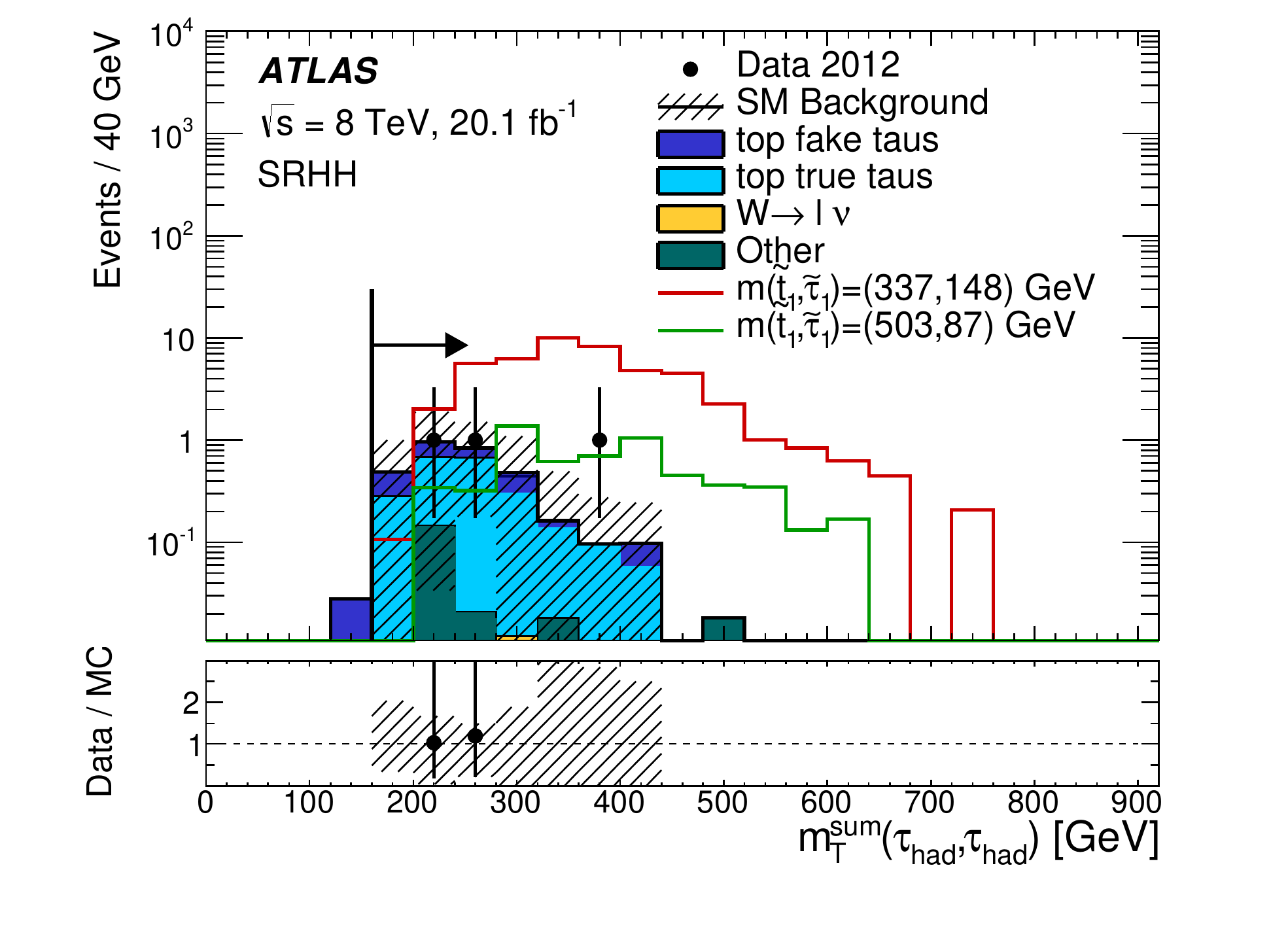} \\
\includegraphics[width=0.80\textwidth]{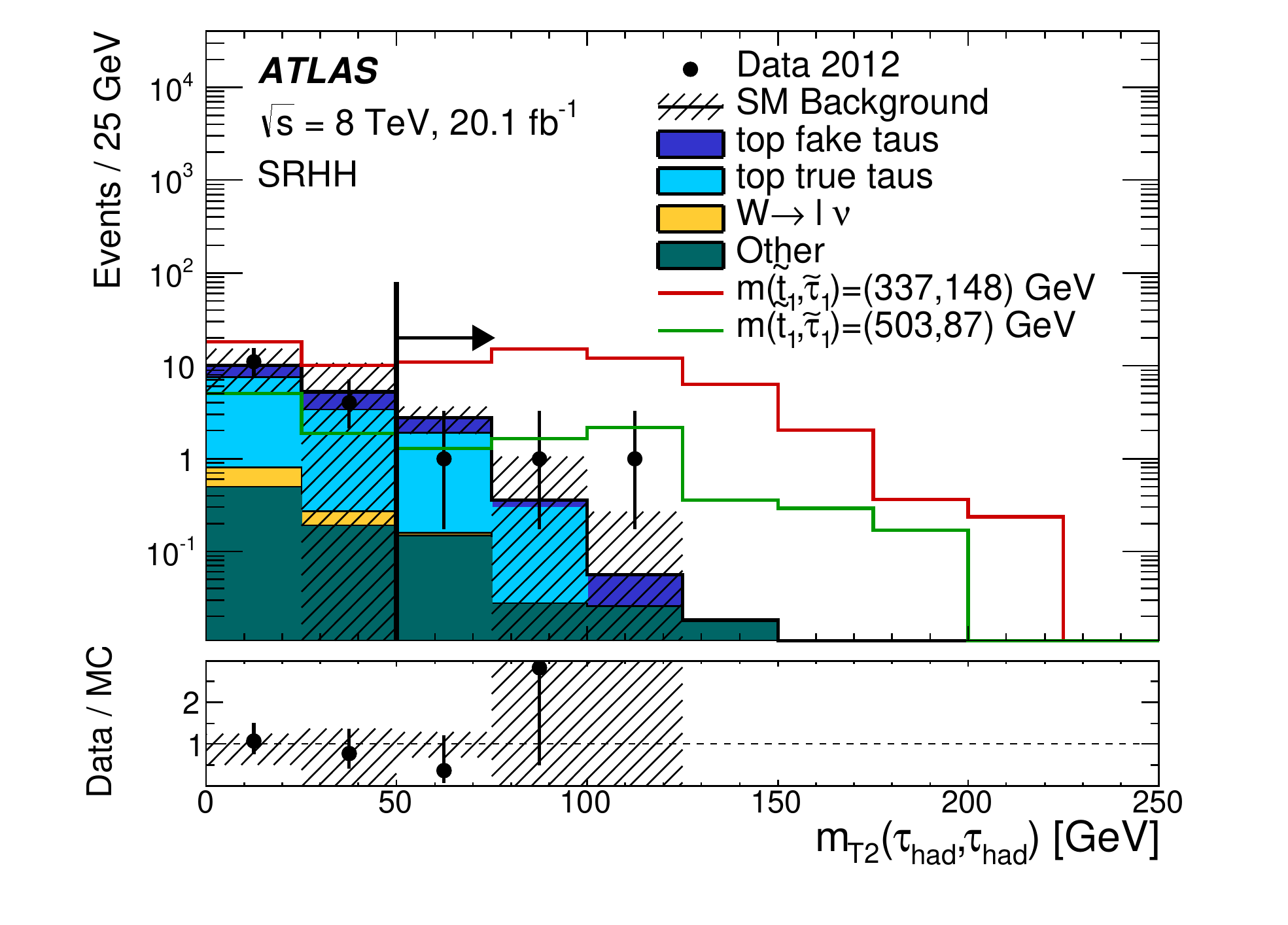}
\end{center}
\caption{Top: distribution of $m_{\rm T}^{\rm sum}(\tauhad,\tauhad)$ for the events passing all the hadron--hadron signal region requirements, except that on $m_{\rm T}^{\rm sum}(\tauhad,\tauhad)$. Bottom: distribution of $m_{{\rm T}2}(\tauhad,\tauhad)$ for the events passing all the hadron--hadron signal region requirements, except that on $m_{{\rm T}2}(\tauhad,\tauhad)$. The contributions from all SM backgrounds are shown as a histogram stack; the bands represent the total uncertainty. The background yields have been rescaled by the post-fit normalisation factors. The arrows mark the cut values used to define the SRs. The distributions expected for two signal models are also shown.} 
\label{fig:SR}
\end{figure}

\begin{figure}[hp!]
\begin{center}
\includegraphics[width=0.80\textwidth]{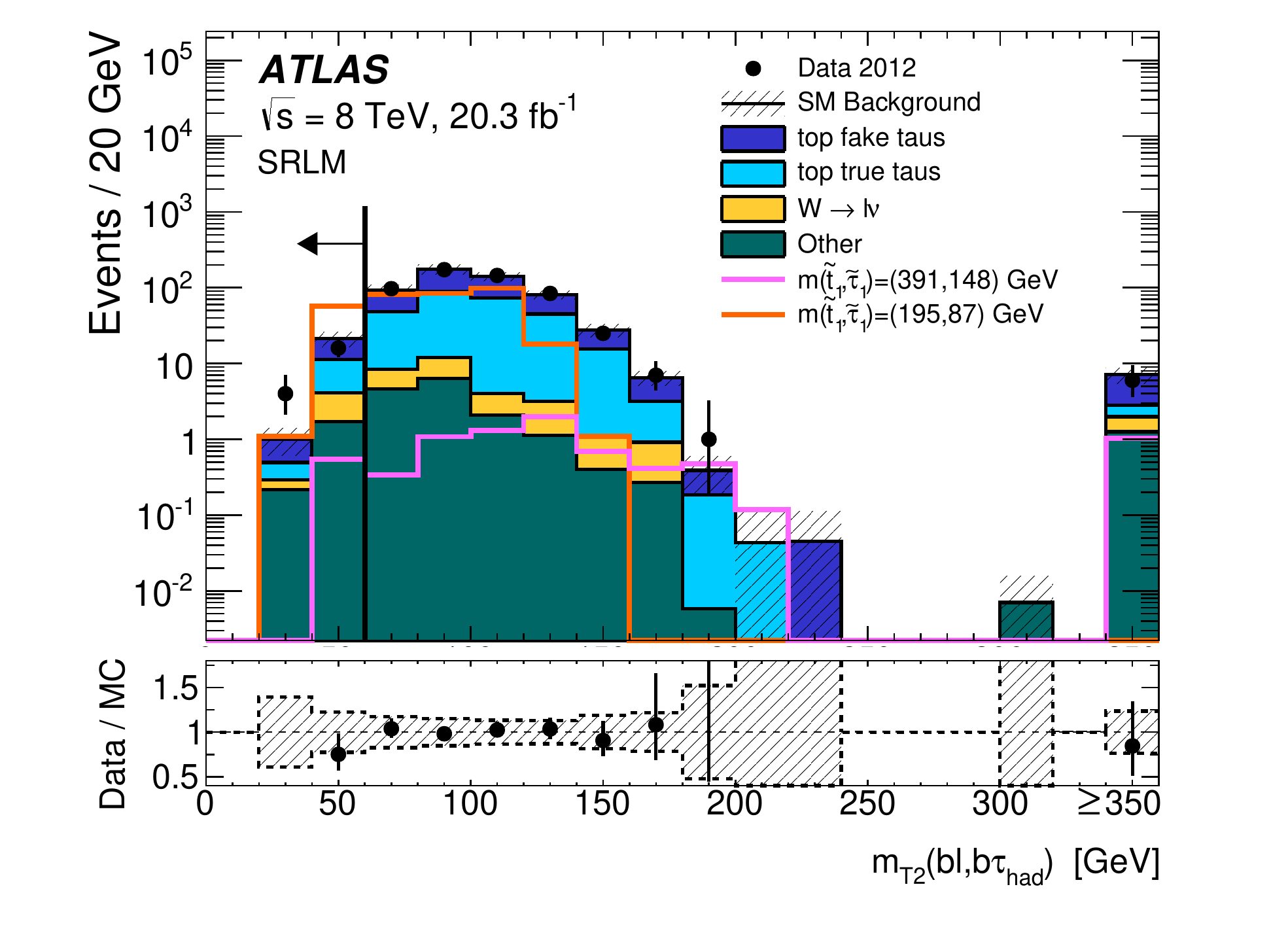} \\
\includegraphics[width=0.80\textwidth]{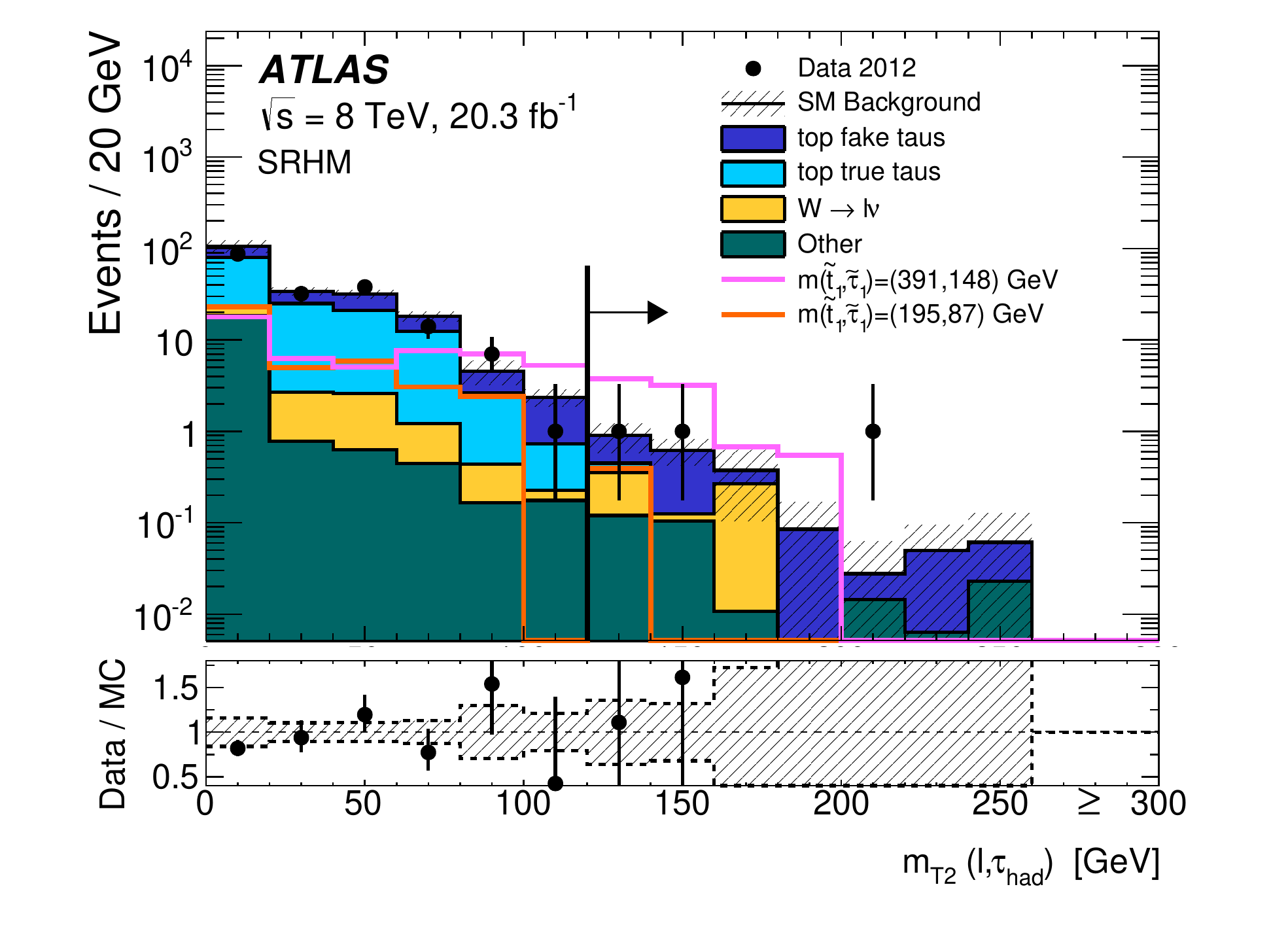}

\end{center}
\caption{Top: distribution of $m_{{\rm T}2}(b \ell, b \tauhad)$ for events passing all the lepton--hadron LM signal region requirements, except that on $m_{{\rm T}2}(b \ell, b \tauhad)$. Bottom: distribution of $m_{{\rm T}2}(\ell, \tauhad)$ for events passing all the lepton--hadron HM signal region requirements, except that on $m_{{\rm T}2}(\ell, \tauhad)$. The contributions from all SM backgrounds are shown as a histogram stack; the bands represent the total uncertainty. The background yields have been rescaled by the post-fit normalisation factors. The arrows mark the cut values used to define the SRs. The overflow bin in the $m_{{\rm T}2}(b \ell, b \tauhad)$\ plot is filled with the events that have for both pairings of $m(b\ell)$ and $m(b\tauhad)$ at least one invariant mass larger than $m_t$. The distributions expected for two signal models are also shown.} 
\label{fig:SR2}
\end{figure}

Upper limits at 95\% confidence level (CL) on the number of beyond-the-SM (BSM) events for each SR are derived using the {\tt HistFitter} program~\cite{HistFitter}, with the CL$_s$ likelihood ratio prescription as described in Ref.~\cite{Read:2002hq}. The limits are calculated for each SR separately, with the observed number of events, the expected background and the background uncertainty as input to the calculation. Possible signal contamination in the control regions is neglected. 
Dividing the limits on the number of BSM events by the integrated luminosity of the data sample, these can be interpreted as upper limits on the visible BSM cross section, $\sigma_{\mathrm{vis}} = \sigma  \times \mathcal{A} \times \epsilon$, where $\sigma$ is the production cross section for the BSM signal, $\mathcal{A}$ is the acceptance defined as the fraction of events passing the geometric and kinematic selections at particle level, and $\epsilon$ is the detector reconstruction, identification and trigger efficiency. 
Table~\ref{tab:upperlimits} summarises, for each SR, the estimated SM background yields, the observed numbers of events, and the expected and observed upper limits on event yields from a BSM signal and on $\sigma_\mathrm{vis}$.  
Table~\ref{tab:acceptanceEfficiency} summarises, for each SR, the acceptance times efficiency for the relevant final state under various signal mass hypotheses.

Exclusion limits are derived for the scalar top pair production, assuming the $\stopOne$ decays with 100\% BR into $b \nu_\tau \stauOne$ and the $\stauOne$ decays into a tau lepton and a gravitino.
The fit used for these limits is similar to that described in Sect.~\ref{sec:selection}, but it now includes the expected signal in the likelihood, with an overall signal-strength parameter constrained to be positive.
The CRs and SRs are fit simultaneously, taking into account the experimental and theoretical systematic uncertainties as nuisance parameters. 
The signal contamination in the CRs is also taken into account.
Exclusion contours are set in the plane defined by the $\stopOne$ and $\stauOne$ masses. 

Systematic uncertainties on the signal expectations stemming from detector effects are included in the fit in the same way as for the backgrounds. Systematic uncertainties on the signal cross section due to the choice of renormalisation and factorisation scales and PDF uncertainties are calculated as described in Sect.~\ref{sec:systematics}. Unlike other nuisance parameters, the signal cross-section uncertainties are only used to assess the impact of a $\pm1\sigma$ variation on the observed limit.

For each mass hypothesis, the expected limits are calculated for the hadron--hadron selection, the two lepton--hadron selections, and the statistical combination of the lepton--lepton selections described in Ref.~\cite{Aad:2014qaa}. The selection giving the best expected sensitivity is used to compute the expected and observed CL$_s$ value. The resulting exclusion contours are shown in Fig.~\ref{fig:limit}. The limits for each individual channel are reported in Fig.~\ref{fig:limit2}. The black dashed and red solid lines show the 95\% CL expected and observed limits, respectively, including all uncertainties except for the theoretical signal cross-section uncertainty (PDF and scale). The yellow bands around the expected limits show the $\pm1\sigma$ expectations. The red dotted $\pm1\sigma$ lines around the observed limit represent the results obtained when varying the nominal signal cross section up or down by its theoretical uncertainty. Numerical limits quoted on the particle masses 
are taken from these $-1\sigma$ {\em theoretical lines}.

As can be seen from Fig.~\ref{fig:limit}, models with a scalar top mass below 490~\GeV\ are excluded.  Depending on the scalar tau mass, some models with scalar top masses up to 650~\GeV\ are also excluded.  The \scalartop\ masses below 150 \GeV\ are not fully considered but they are unlikely to be viable because the cross section
times branching ratio for $\stopOne \stopOne \rightarrow b \tau b \tau + X$\ is more than 25 times larger than the cross section times branching ratio for the
production of \ttbar\ decaying into the same di-tau final state, and measurements of the \ttbar\ cross section in various final states \cite{TOPQ-2011-05, CMS:ttbarCrossSection1, TOPQ-2012-04, CMS:ttbarCrossSection2}\ are in good agreement with the SM prediction.

\begin{figure}[!hbp]
\begin{center}
\includegraphics[width=0.8\textwidth]{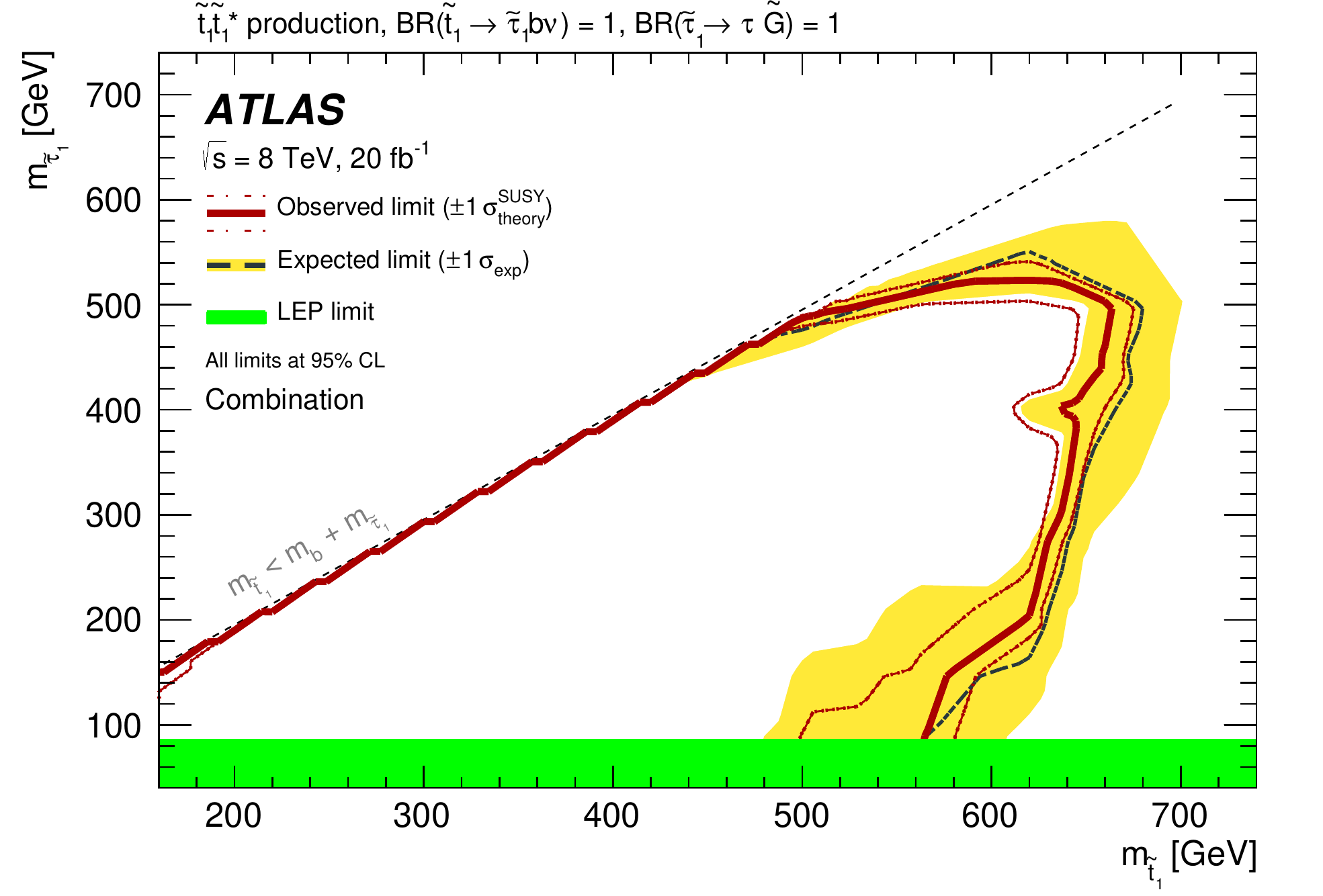}

\end{center}
\caption{\label{fig:limit} Observed and expected exclusion contours at 95\% CL in the ($\stopOne,\stauOne$) mass plane
from the combination of all selections.  The dashed and solid lines show the 95\% CL expected and observed limits, respectively, including all uncertainties except for the theoretical signal cross-section uncertainty (PDF and scale). The band around the expected limit shows the $\pm1\sigma$ expectation. The dotted $\pm1\sigma$ lines around the observed limit represent the results obtained when varying the nominal signal cross section up or down by the theoretical uncertainty.
The LEP limit on the mass of the scalar tau is also shown.}
\end{figure}

\begin{figure}[!hbp]
\begin{center}
\includegraphics[width=7cm,height=5cm]{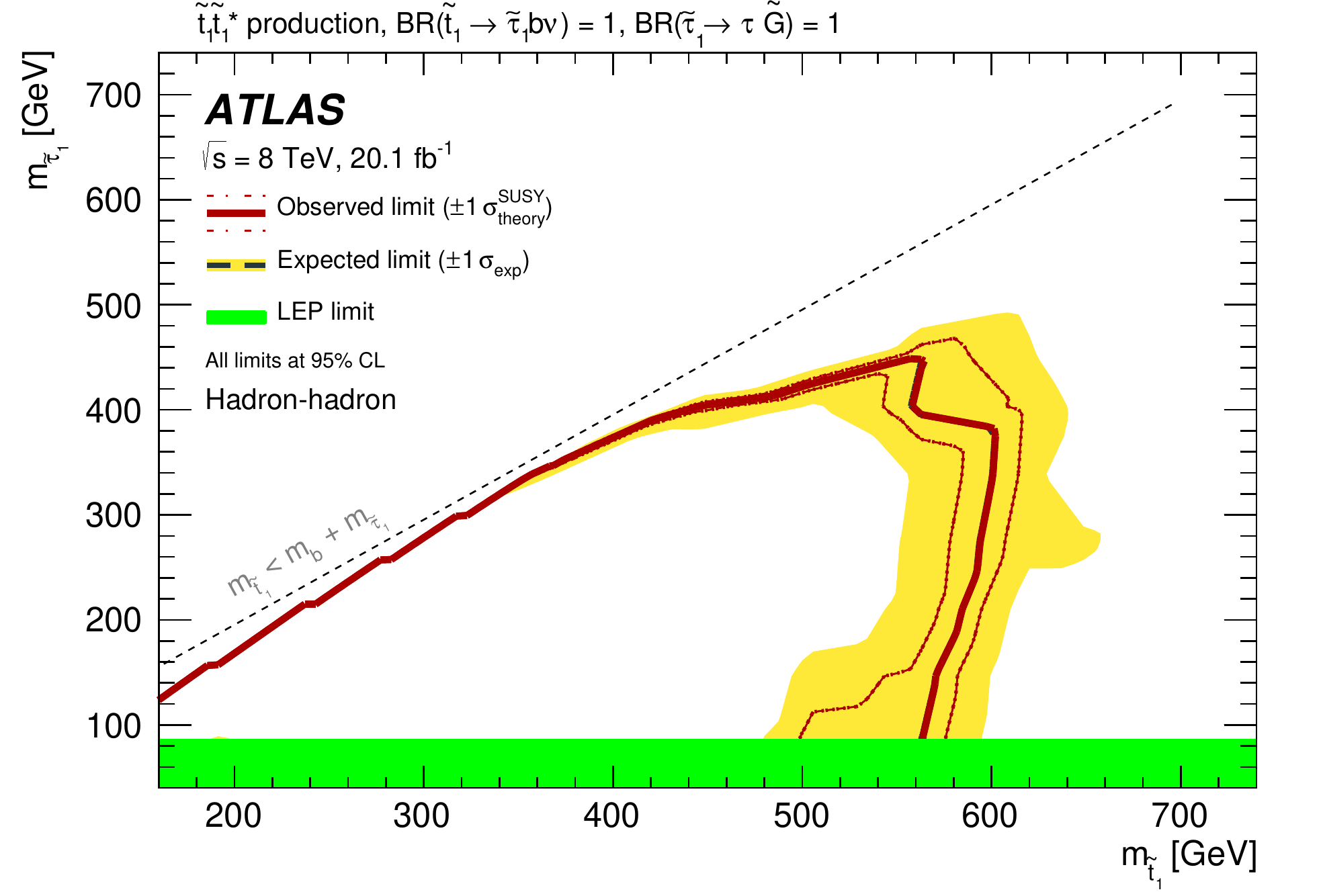} 
\includegraphics[width=7cm,height=5cm]{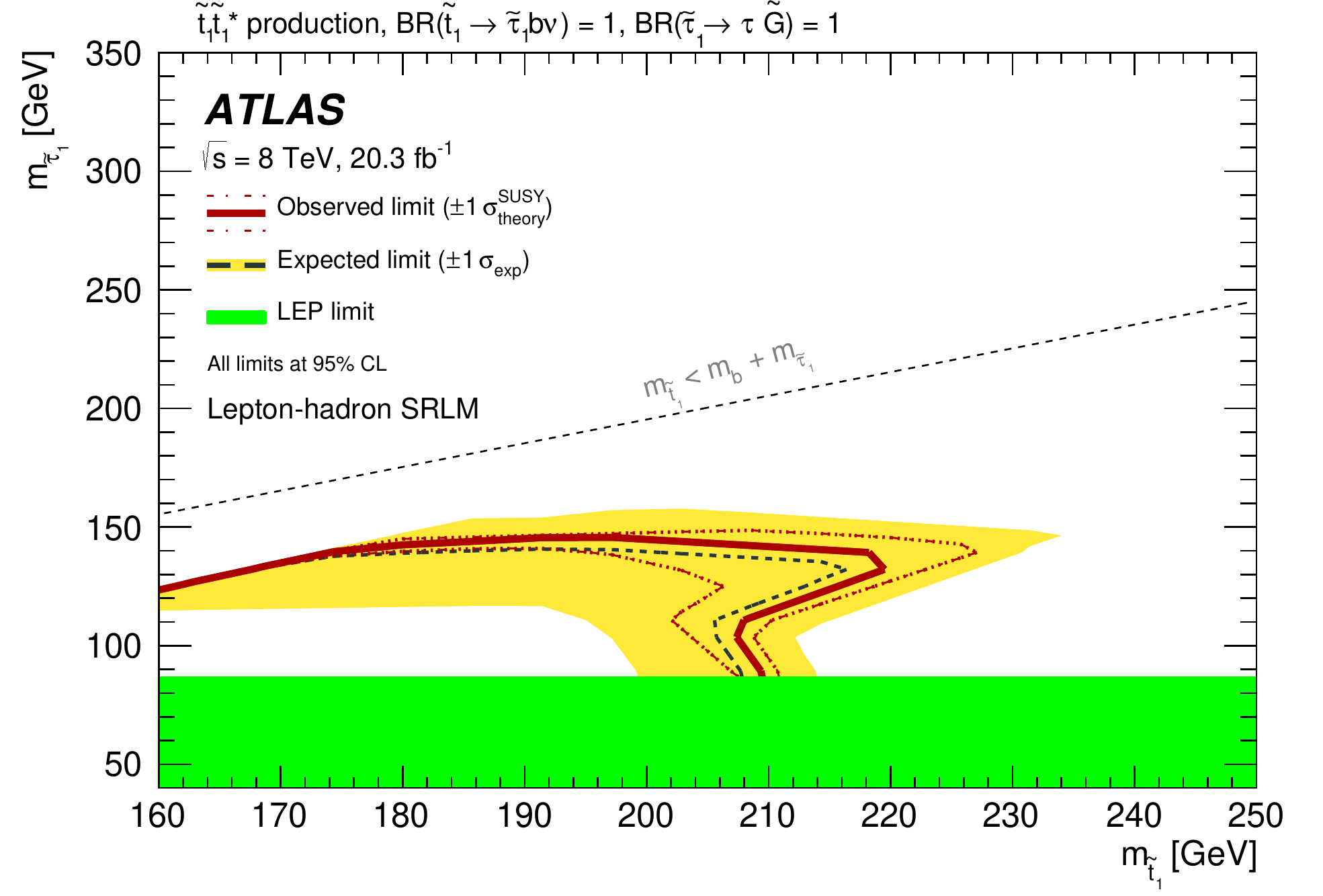}  \\
\includegraphics[width=7cm,height=5cm]{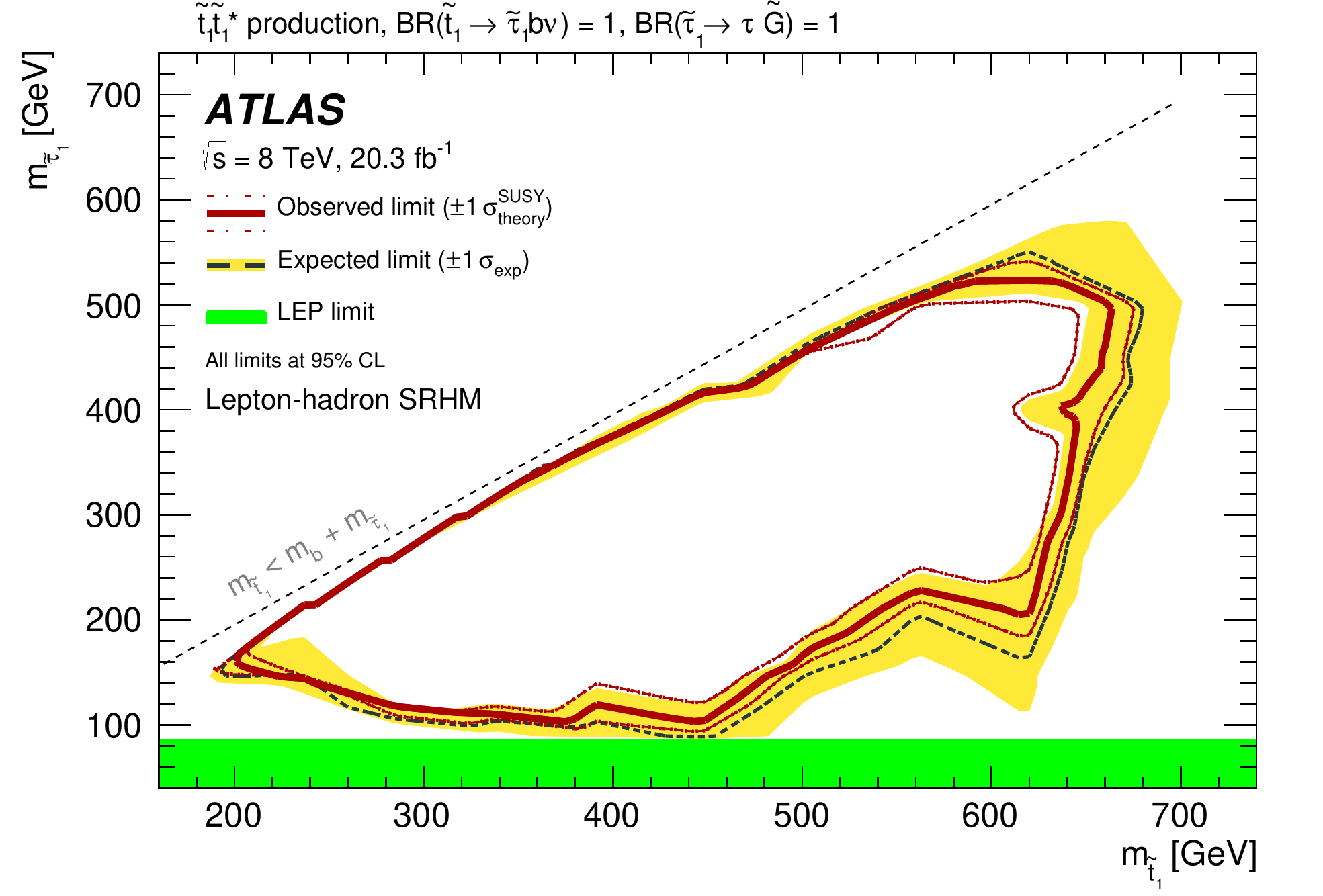}
\includegraphics[width=7cm,height=5cm]{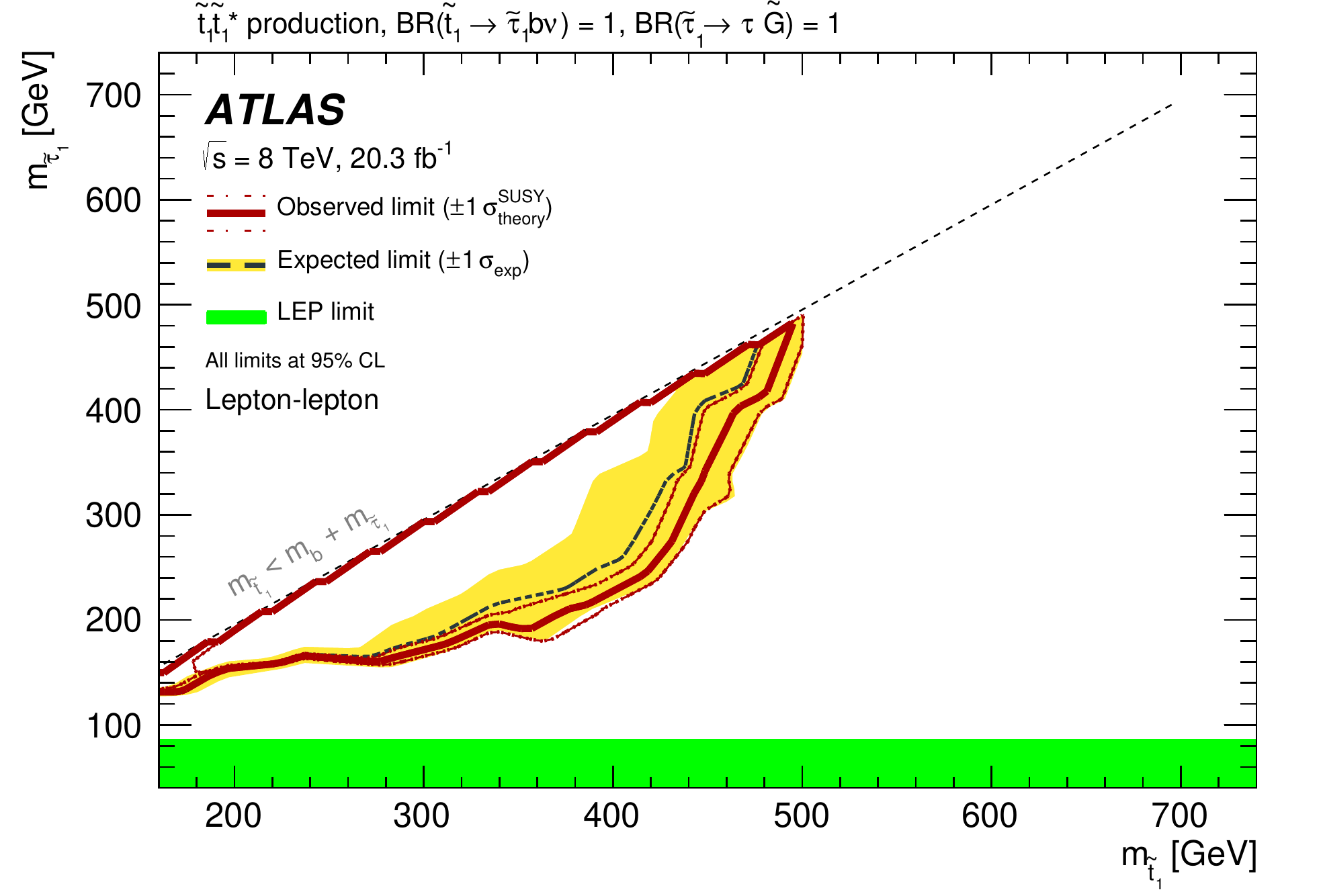} \\

\end{center}
\caption{\label{fig:limit2} Observed and expected exclusion contours at 95\% CL in the ($\stopOne, \stauOne$) mass plane
from the hadron--hadron (top left), the lepton--hadron low-mass (top right), the lepton--hadron high-mass (bottom left) and the 
lepton--lepton selections of Ref.~\cite{Aad:2014qaa} (bottom right).  The dashed and solid lines show the 95\% CL expected and observed limits, respectively, including all uncertainties except for the theoretical signal cross-section uncertainty (PDF and scale). The band around the expected limit shows the $\pm1\sigma$ expectation. The dotted $\pm1\sigma$ lines around the observed limit represent the results obtained when varying the nominal signal cross section up or down by the theoretical uncertainty.
The LEP limit on the mass of the scalar tau is also shown.}
\end{figure}

\newpage
\section{Conclusion}
\label{sec:conclusion}

A search for direct pair production of supersymmetric partners of the top quark decaying via a scalar tau to a nearly massless gravitino
has been performed using  20~fb$^{-1}$ of $pp$ collision data at $\sqrt{s}=8$~\TeV, collected by the ATLAS experiment at the LHC in 2012. 
Scalar top candidates are searched for in events with either two hadronically decaying taus, one hadronically decaying tau and one light lepton, or two light leptons.
Good agreement is observed between the Standard Model background estimate and the data. 
The first results from a hadron collider search for the three-body decay mode to the scalar tau are presented.  In the context of the model considered, lower limits on the scalar top mass are set at 95\% confidence level, and found to be between 490~\GeV\ and 650~\GeV\ for scalar tau masses ranging from the LEP limit to the scalar top mass. 

\section*{Acknowledgements}

\input{acknowledgements/Acknowledgements}

\clearpage

\printbibliography

\newpage \input{atlas_authlist}

\end{document}

%% file: paper-metadata.tex
\AtlasTitle{Search for direct top squark pair production in final states with two tau leptons in $pp$ collisions at $\sqrt{s}=8$~\TeV\ with the ATLAS detector}

\usepackage{authblk}
\renewcommand\Authands{, } 
\renewcommand\Affilfont{\itshape\small} 

\usepackage{authblk}

\AtlasRefCode{SUSY-2014-04}

\PreprintIdNumber{CERN-PH-EP-2015-188}

\AtlasAbstract{  A search for direct pair production of the supersymmetric partner of the top quark, decaying via a scalar tau to a nearly massless gravitino, has been performed using  20~fb$^{-1}$ of proton--proton collision data at $\sqrt{s}=8 \TeV$. The data were collected by the ATLAS experiment at the LHC in 2012. Top squark candidates are searched for in events with either two hadronically decaying tau leptons, one hadronically decaying tau and one light lepton, or two light leptons. No significant excess over the Standard Model expectation is found.  Exclusion limits at $95\%$~confidence level are set as a function of the top squark and scalar tau masses. Depending on the scalar tau mass, ranging from the $87 \GeV$~LEP limit to the top squark mass, lower limits between $490 \GeV$~and $650 \GeV$~are placed on the top squark mass within the model considered.}

\AtlasCoverCommentsDeadline{20 July 2015}
\AtlasJournal{EPJC}

\AtlasCoverAnalysisTeam{Maria Ilaria Besana, Pierfrancesco Butti, Ewan Hill, Tommaso Lari~(*), Bertrand Martin dit Latour, Chiara Rizzi, Fabrizio Salvatore, Isabel Marian Trigger, Sarah Williams}

\AtlasCoverEdBoardMember{Lucio Cerrito~(*)}
\AtlasCoverEdBoardMember{Matthew Beckingham}
\AtlasCoverEdBoardMember{Ulla Blumenschein}
\AtlasCoverEdBoardMember{Mark Kruse}

\AtlasCoverEgroupEditors{atlas-SUSY-2014-04-editors@cern.ch}

\AtlasCoverEgroupEdBoard{atlas-SUSY-2014-04-editorial-board@cern.ch}

%% file: acknowledgements/Acknowledgements.tex

We thank CERN for the very successful operation of the LHC, as well as the
support staff from our institutions without whom ATLAS could not be
operated efficiently.

We acknowledge the support of ANPCyT, Argentina; YerPhI, Armenia; ARC,
Australia; BMWFW and FWF, Austria; ANAS, Azerbaijan; SSTC, Belarus; CNPq and FAPESP,
Brazil; NSERC, NRC and CFI, Canada; CERN; CONICYT, Chile; CAS, MOST and NSFC,
China; COLCIENCIAS, Colombia; MSMT CR, MPO CR and VSC CR, Czech Republic;
DNRF, DNSRC and Lundbeck Foundation, Denmark; EPLANET, ERC and NSRF, European Union;
IN2P3-CNRS, CEA-DSM/IRFU, France; GNSF, Georgia; BMBF, DFG, HGF, MPG and AvH
Foundation, Germany; GSRT and NSRF, Greece; RGC, Hong Kong SAR, China; ISF, MINERVA, GIF, I-CORE and Benoziyo Center, Israel; INFN, Italy; MEXT and JSPS, Japan; CNRST, Morocco; FOM and NWO, Netherlands; BRF and RCN, Norway; MNiSW and NCN, Poland; GRICES and FCT, Portugal; MNE/IFA, Romania; MES of Russia and ROSATOM, Russian Federation; JINR; MSTD,
Serbia; MSSR, Slovakia; ARRS and MIZ\v{S}, Slovenia; DST/NRF, South Africa;
MINECO, Spain; SRC and Wallenberg Foundation, Sweden; SER, SNSF and Cantons of
Bern and Geneva, Switzerland; NSC, Taiwan; TAEK, Turkey; STFC, the Royal
Society and Leverhulme Trust, United Kingdom; DOE and NSF, United States of
America.

The crucial computing support from all WLCG partners is acknowledged
gratefully, in particular from CERN and the ATLAS Tier-1 facilities at
TRIUMF (Canada), NDGF (Denmark, Norway, Sweden), CC-IN2P3 (France),
KIT/GridKA (Germany), INFN-CNAF (Italy), NL-T1 (Netherlands), PIC (Spain),
ASGC (Taiwan), RAL (UK) and BNL (USA) and in the Tier-2 facilities
worldwide.

%% file: atlas_authlist.tex
\begin{flushleft}
{\Large The ATLAS Collaboration}

\bigskip

G.~Aad$^{\rm 85}$,
B.~Abbott$^{\rm 113}$,
J.~Abdallah$^{\rm 151}$,
O.~Abdinov$^{\rm 11}$,
R.~Aben$^{\rm 107}$,
M.~Abolins$^{\rm 90}$,
O.S.~AbouZeid$^{\rm 158}$,
H.~Abramowicz$^{\rm 153}$,
H.~Abreu$^{\rm 152}$,
R.~Abreu$^{\rm 116}$
% ...
\bigskip
\\
% ...
$^{11}$ Institute of Physics, Azerbaijan Academy of Sciences, Baku, Azerbaijan\\
% ...
$^{85}$ CPPM, Aix-Marseille Universit{\'e} and CNRS/IN2P3, Marseille, France\\
% ...
$^{90}$ Department of Physics and Astronomy, Michigan State University, East Lansing MI, United States of America\\
% ...
$^{107}$ Nikhef National Institute for Subatomic Physics and University of Amsterdam, Amsterdam, Netherlands\\
% ...
$^{113}$ Homer L. Dodge Department of Physics and Astronomy, University of Oklahoma, Norman OK, United States of America\\
% ...
$^{116}$ Center for High Energy Physics, University of Oregon, Eugene OR, United States of America\\
% ...
$^{151}$ Institute of Physics, Academia Sinica, Taipei, Taiwan\\
$^{152}$ Department of Physics, Technion: Israel Institute of Technology, Haifa, Israel\\
$^{153}$ Raymond and Beverly Sackler School of Physics and Astronomy, Tel Aviv University, Tel Aviv, Israel\\
% ...
$^{158}$ Department of Physics, University of Toronto, Toronto ON, Canada\\
% ...
\end{flushleft}
